\documentclass[preprint,tightenlines, aps,prd,floats,nofootinbib,amssymb,groupedaddress]{revtex4-1}
\usepackage[sort&compress]{natbib}
\usepackage{epsf,epsfig}
\usepackage{amsmath}
\usepackage{hyperref}
\usepackage{subfigure}
\usepackage{rotating, graphicx}

\usepackage[dvipsnames]{xcolor}
\usepackage{color,colortbl}

\usepackage{mathrsfs}
\usepackage{pgf,tikz,tikz-feynman}

\newcommand{\be}{\begin{equation}}
\newcommand{\ee}{\end{equation}}
\newcommand{\ba}{\begin{array}}
\newcommand{\eaq}{\end{array}}
\newcommand{\bea}{\begin{eqnarray}}
\newcommand{\eea}{\end{eqnarray}}

\newcommand{\nn}{\nonumber}

\newcommand{\bi}{\begin{itemize}}
\newcommand{\ei}{\end{itemize}}
\newcommand{\bal}{\begin{aligned}}
\newcommand{\eal}{\end{aligned}}
\newcommand{\Tr}{\operatorname{Tr}}
\newcommand{\Exp}{\operatorname{Exp}}

\usepackage{comment,booktabs}
\usepackage{wrapfig}

\begin{document}
\title{Effects of bulk  symmetry breaking on AdS/QCD predictions}
\author{Dom\`enec Espriu}
\email{espriu@icc.ub.edu}
\author{Alisa Katanaeva}
\email{katanaeva@fqa.ub.edu}
\affiliation{\it Departament de F\'\i sica Qu\`antica i Astrof\'\i sica and \\
Institut de Ci\`encies del Cosmos (ICCUB), Universitat de Barcelona,\\ 
Mart\'i i Franqu\`es 1, 08028 Barcelona, Catalonia, Spain}
\begin{abstract}

We put forward a new bottom-up AdS/QCD holographic model bearing a distinct treatment of the pion fields. We argue that a standard approach to the pion description is neither transparent nor totally satisfactory. In the paper we provide a new one based on a broadened realization of some holographic principles. The reasoning and the effect of these modifications are explained in detail. The resulting model has a different set of parameters than the standard AdS/QCD case. We use them to calculate an extensive list of QCD quantities and find a rather good agreement with the experimental data.

\end{abstract}
\maketitle

\section{Introduction}
A wide range of studies in the bottom-up AdS/QCD holography is devoted to the five-dimensional dual description of the light vector and scalar mesons in association with a realization of the chiral symmetry and the products of its breaking. The first to appear were the Hard Wall (HW)~\cite{HW_2005, daRold_2005,Hirn2005} and the Soft Wall (SW)~\cite{SW_2006} models. The two utilize  conceptually distinct ways of implementing the bulk Lagrangian in the  five-dimensional anti-de Sitter (AdS) space-time. In these simplest setups it was attempted to describe the phenomenology of the vector sector and its interaction with the pions in Refs.~\cite{HW_2005, daRold_2005,Hirn2005,SW_2006,Grigoryan2007} and  the scalar sector separately was considered in Ref.~\cite{DaRold_scalar,Colangelo2008}.
The mentioned models also differ in the way the chiral symmetry breaking is introduced. Various modifications and extensions followed, and the ones relevant to this paper will be mentioned in the text.

In order to build a 5D model within the bottom-up approach one mixes the established AdS/CFT prescriptions with various assumptions. The latter might have a theoretical motivation, but the ultimate criterion for their validity is leading to a better phenomenological description for one or another aspect of QCD. The most prominent example is the introduction of some kind of ``wall'' to break the conformal invariance of the AdS metric. The proposals of a sharp cut-off (HW), a smooth exponential one (SW), or something in between  (e.g., \cite{Kwee2008,Gherghetta2009}) compete on phenomenological grounds. 

This said, we strongly believe that the field of viable model modifications is not exhausted yet. In this paper we construct and investigate a new holographic framework that is based on the SW  setup and is dual to $SU(2)$ QCD. From a theoretical point of view, our goal is a new consistent description of the Goldstone states (pions).
In the common holographic setup the Goldstone bosons turn out to be a part of the gauge field (playing the role of the "Higgs"). This is not the way chiral symmetry is broken in real QCD. One way around this would be introducing some symmetry breaking terms: $5D$ vector meson masses in order to prevent Goldstones from being eaten, and an extra scalar term to make the Goldstone masses ($m_\pi$) lower than the natural scale of the composite states ($m_\rho$) and to fulfill the holographic requirements on the profile of the relevant mode in the extra dimension. Furthermore, by doing so we would be able to treat the Goldstones in a transparent and analytically tractable fashion that is often lacking in other approaches.

There is also an issue on the phenomenological side. Various low-energy observables have been calculated within one or another model and are claimed to be in agreement with experiment at level of $10-30\%$, but it is rather common that a given study is concentrated on a specific set of observables.
The purpose of this work is to be as exhaustive as possible and to make as many predictions for the observables as can be extracted from this particular model of QCD with two flavors up to the three-point level.

In the process we realized that re-estimation and generalization of some concepts of the holographic model construction are necessary. The first one concerns the duality between the QCD operator and the five-dimensional field. Another concerns the mass prescriptions for these fields. Both are established in the so-called AdS/CFT dictionary \cite{Maldacena_1999,*Gubser1998,*Witten_1998,*Klebanov_1999}, but we dispute its blind following in the phenomenology directed approach of AdS/QCD. 

The dual operators in the dictionary are understood rather abstractly, for once they have no fixed normalizations attached. We suggest introducing some reference operators with free coefficients and studying whether they are eliminated from the physical quantities or not. The holographic prescriptions for the $5D$ masses in the dictionary are extremely stringent to the model. We argue that they should rather be considered as imposing boundary conditions on otherwise bulk coordinate-dependent mass (not the first attempt on this, see e.g. \cite{Forkel_2007,Vega2010,Fang2016}). 
Obviously, a non-zero vector mass means that the local symmetry is not preserved in the holographic action in the bulk, but we will see that it is kept on the boundary. In addition to this, we also introduce an explicit breaking of the global chiral symmetry towards the vector subgroup in the scalar sector. That is not conventional but it turns out that this kind of symmetry breaking is crucial to achieving our goal regarding pions.
 
 We would like to stress that the cumulative effect of all these modifications of the standard bottom-up framework turns out to be more interesting than was pre-designed. 
 For instance, just demanding the analyticity of solutions of the equations of motion results in a determined ansatz for the scalar vacuum expectation value (vev), which is the driver behind the chiral symmetry breaking in the holographic bulk. 
 That leads us to question the common parametrization of the scalar vev in terms of the quark mass and chiral condensate (see also \cite{Erlich2008,Gherghetta2009,Cherman2009}).
 The two aforementioned non-standard symmetry violations and this particular choice of the scalar vev determine the novel phenomenology of our model.
Besides, the appearance of several new parameters hints for a better fit to experiment. Moreover, after a close examination we will find out that the number of free parameters could be minimized to that of the traditional SW, while the described phenomenology remains richer.

 The structure of this work is as follows. In Section~\ref{5Dmodel} we describe the way the model emerges, starting with the set of relevant QCD operators and showing the elaborated evolution from the standard approach. In Section~\ref{5DEOM} the solutions to the holographic equations of motion are obtained. The Regge trajectories of the radial excitations of $\rho$, $a_1$, $a_0$, $\pi$ mesons (the linearity of which is guaranteed by the SW) together with their decay constants are the phenomenological quantities of interest there. The structure of the two-point correlators is well studied in QCD, their high energy limit analyzed thanks to the operator product expansion (OPE). We provide the holographic results for them in Section~\ref{2ptCorrelator}. The three-point correlation functions of Section~\ref{3ptCorrelator} give a lot of information on several coupling constants and the form factors. Finally, in Section~\ref{Experiment} we summarize all the observables considered and make some global and particular fits to determine the model parameters. We conclude in Section~\ref{Conclusions}.

\section{Holographic model}\label{5Dmodel}
\subsection{Connection to the $4D$ QCD}
The construction of the 5D model begins by selecting a collection of operators describing the 4D physics of interest. We use a standard set of QCD operators representative of the chiral flavor symmetry and its breaking.

We focus on the two flavor case and work with the $SU(2)$ generators  $T_a=\sigma_a/2$, such that $[T_a,T_b]=i\varepsilon_{abc} T_c,\ \Tr(T_a T_b)=\delta_{ab}/2$, $a,b=1,2,3$.
      
In the vector sector the conserved QCD currents are 
$\mathcal O^{a\ \mu}_L=\overline\Psi_L\gamma^\mu (T^{a})\Psi_L,$ and
$\mathcal O^{a\ \mu}_R=\overline\Psi_R\gamma^\mu (T^{a})\Psi_R$.
Their combinations result in
the vector operator $\mathcal O^{a\ \mu}_V=\mathcal O^{a\ \mu}_L+\mathcal O^{a\ \mu}_R$ and the axial vector operator $\mathcal O^{a\ \mu}_A=\mathcal O^{a\ \mu}_R-\mathcal O^{a\ \mu}_L$.
The scalar condensate transforms as 
$\langle\overline\Psi_R\Psi_{L}\rangle=({\bf 2}, {\bf \overline 2})_{SU(2)_L\times SU(2)_R}$ and produces 
the breaking pattern $SU(2)_L\times SU(2)_R \rightarrow SU(2)_V$.
The scalar bilinears are given in terms of the flavor components of $\Psi$:
$\overline\Psi_R^{j}\Psi_{L}^k$ and its conjugate $\overline\Psi_L^{j}\Psi_{R}^k$.
We note that these QCD operators have some specific normalization, which we shall keep as a reference one. 

Within the holographic approach the consideration of the partition function $\mathcal Z_{4D}$ is the cornerstone concept. Its conventional structure is
\be\label{partF}
\mathcal Z_{4D}[\phi_{\mathcal O}]=\int[\mathcal D \Psi][\mathcal D \overline\Psi] \Exp i\int d^4x [\mathcal L_{QCD}(x) + \sum_j\phi_{\mathcal O_j}(x)\mathcal O_j(x)],
\ee
where $\phi_{\mathcal O}$ are the sources of the corresponding operators.
 In holography one relies on an assumption that the very functional with the integration over the fundamental degrees of freedom performed can be associated with a quantity derived from the $5D$ by reducing the extra dimension \cite{Maldacena_1999,*Gubser1998,*Witten_1998,*Klebanov_1999}. To exploit the holographic procedure there is no necessity to talk about a particular normalization of a given operator; in the dictionary they are differentiated just by their canonical dimension and spin.  However, some phenomenological observables in QCD may turn out to be dependent on the normalization. Thus, to see the possible impact of the normalization choice, we introduce extra factors $g_V$ and $g_S$ in the vector and scalar operators, respectively. 
 
At the same time, we would like to couple the QCD currents to the electroweak bosons of the SM. There the symmetry leaves no ambiguity for the couplings given in terms of the electroweak coupling constants $e$ and $g$.

 The breaking related operator is a bilinear, and hence its source is a matrix. We can make the following interpretation of it:
 $\phi_{\overline{\Psi} \Psi}=m_q\cdot \text{Id}+\phi^a_S\cdot T^a-i\phi^a_P\cdot T^a$, where $m_q$ is a physical source related to the quark mass.
The other two non-physical sources in the expansion imply that we can introduce a proper scalar $ \mathcal  O_S^a=(T^a)_{jk}\left(\overline\Psi_R^{j}\Psi_{L}^k+\overline\Psi_L^{j}\Psi_{R}^k\right),$ and pseudoscalar operator $\mathcal O_P^a=i(T^a)_{jk}\left(\overline\Psi_L^{j}\Psi_{R}^k-\overline\Psi_R^{j}\Psi_{L}^k\right)$.
 
To conclude, in the partition function~(\ref{partF}) the relevant QCD operators appear as follows in our setup
\bea
\sum_j\phi_{\mathcal O_j}(x)\mathcal O_j(x) &=& \phi_V^{a\ \mu}(x)\cdot g_V \mathcal O_{V\ \mu}^a(x)+
\phi_A^{a\ \mu}(x)\cdot g_V \mathcal O_{A\ \mu}^a(x)\\ 
&&+ \phi^a_{S}(x)\cdot g_S  \mathcal O_S^a(x) +  \phi^a_{P}(x)\cdot g_S  \mathcal O_P^a(x)\nn\\
&&+e \mathcal A^{em}_\mu\cdot \mathcal O_{V}^{3\ \mu} -\frac g2 W^{-/+}_\mu\cdot \mathcal O_{A}^{-/+\ \mu}+\frac g2  W^{-/+}_\mu\cdot  \mathcal O_{V}^{-/+\ \mu}+...\nn
\eea
 where we use the notation $\mathcal O^+=\frac{\mathcal O^1+i\mathcal O^2}{\sqrt2}$, $\mathcal O^-=\frac{\mathcal O^1-i\mathcal O^2}{\sqrt2}$. 

\subsection{Standard $5D$ construction}
Applying the gauge-gravity correspondence to the aforementioned operators we obtain a theory for the left and right vector fields and a complex scalar field. The holographic dictionary provides relations between operators and $5D$ fields and dictates the masses of the latter:
\bea
&g_V \mathcal O^{a}_{L\ \mu}\leftrightarrow (A_L)^a_\mu,\quad g_V \mathcal O^{a}_{R\ \mu}\leftrightarrow (A_R)^a_\mu,\quad &M^2_LR^2=M^2_RR^2=0;\\
&g_S \overline\Psi_R^{j}\Psi_{L}^k \leftrightarrow \frac Rz H^{jk},\ g_S \overline\Psi_L^{j}\Psi_{R}^k \leftrightarrow \frac Rz H^{\dagger j k},\quad &M^2_HR^2=-3.
\eea

Matter fields live in a curved five-dimensional AdS space of radius $R$  with the metric $$g_{MN}=\frac{R^2}{z^2}\eta_{MN},\qquad \eta_{MN}=\text{diag}\{1,-1,-1,-1,-1\}.$$

The global symmetries of QCD translate into the local ones on the $5D$ side.
Consideration of the transformation properties of different fields allows us to construct a gauge invariant Lagrangian with spontaneous symmetry breaking to the diagonal (vector) subgroup, $SU(2)_L\times SU(2)_R \rightarrow SU(2)_V$. 

Let us denote the group transformations $g_L\in SU(2)_L,\ g_R\in SU(2)_R,\ h\in SU(2)_V$.
The canonical choice for the coset representative $\xi(\pi)=(\xi_L(\pi),\xi_R(\pi))\in SU(2)_L\times SU(2)_R$ is to take $\xi_L=\xi_R^\dagger=u(\pi)$. Then the matrix of the Goldstone fields  goes as follows under a chiral transformation: $u\rightarrow u'=g_Luh^\dagger=hug_R^\dagger$. The scalar degrees of freedom are collected in $\Sigma$ transforming as $\Sigma\rightarrow\Sigma'=h\Sigma h^\dagger$.
With these we construct a non-linear complex scalar field $H(x,z)$
\be
H=u\Sigma u,\ \Sigma=f(z)\cdot \text{Id}+ T^as^a(x,z),\ u=\exp\left(\frac{i\pi^a(x,z) T^a}{\chi_\pi}\right),
\ee
for which we have $H\rightarrow H'=g_LHg_R^\dagger$. $\chi_\pi$ is a constant parameter used to normalize the dimensionality of the $\pi$ fields. There is no reason to immediately connect it to the QCD pion decay constant, the commonly used scale. The scalar vev, $f(z)$, implements chiral symmetry breaking in the bulk. This will be discussed in more detail further.

In the vector sector we have the non-Abelian fields $(A_L)_M$ and $(A_R)_M $, their kinetic terms given by the field strength tensor $F_{MN}=(\partial_MA_N^a-\partial_NA_M^a+C^{abc}A^b_MA^c_N)T^a$. The covariant derivative transforming as $D_MH\rightarrow g_LD_MHg_R^\dagger$ is 
\be
D_MH=\partial_M H-iA_{LM}H+iHA_{RM}.
\ee

The general dynamics is governed by the $5D$ action:
\bea
S_{5D}&=&-\frac1{4g_5^2}\int d^5x\sqrt{-g}e^{-\Phi(z)}\Tr\left[F^L_{MN}F^{L\ MN}+F^R_{MN}F^{R\ MN}\right]\\
&&+\frac1{k_s}\int d^5x\sqrt{-g}e^{-\Phi(z)}\left[ \Tr g^{MN}(D_MH)^\dagger(D_NH)
-M^2_H \Tr HH^\dagger \right]. \nn
\eea
We introduce here the holographic parameters $[g_5^2]=[k_s]=E^{-1}$ in order to retain the standard dimensionalities of the fields.
The particular holographic model is also determined by the SW setup implemented through the dilaton profile $\Phi(z)=\kappa^2 z^2$, where $\kappa$ is a model parameter setting an overall energy scale  \cite{SW_2006}.

\subsection{Symmetry breaking in the bulk}
The major disadvantage of the standard construction, from our point of view, is that pions, being introduced as they are, appear at the two-point level just in a combination $(\partial_M \pi-A_M)^2$. That makes them quite similar to the Goldstones in the Higgs mechanism and wrongly implies that they are fully dedicated to contribute to the axial two-point function (analogous to the mass of a gauge boson). It is known that the QCD pion should do more than that.

We want to make some changes in the setup so that the pion can no longer be eliminated by the gauge choice. The proposal consists in the introduction of a term providing a non-trivial diagonalization on $(A_M,\partial_M \pi)$ plane. The natural option is to add some $z$-dependence to the masses dictated by the holographic dictionary. Other authors \cite{Forkel_2007,Fang2016} have looked into this option motivated by a different reasoning, and the focus usually stays on the scalar mass \cite{Paula2010,Vega2010,Fang2016} on the grounds that its $z$-dependent part could be attributed to the anomalous dimension of the relevant quark operator. Obviously, by including masses for the $5D$ gauge fields, we give up the local chiral gauge invariance. The following expressions for the vector and scalar masses will be used in this work:
\bea
M^2_LR^2=M^2_RR^2=M^2(z)R^2=0+4\mu_V \kappa^2z^2,\\
M_H^2(z)R^2=-3+4\mu_H \kappa^2z^2.
\eea
The quadratic in $z$ terms with $\mu_V$ and $\mu_H$ represent a minimal option to achieve the stated purpose while keeping the solutions analytically tractable.

For reasons that shall become clear further on we also include a  scalar potential term containing a new function $b(z)$, that explicitly breaks the axial part of the symmetry.
The total five-dimensional action of our model will be
\bea
&S=-\frac1{4g_5^2}\int d^5x\sqrt{-g}e^{-\Phi(z)}\Tr\left[F^L_{MN}F^{L\ MN}+F^R_{MN}F^{R\ MN}-2M^2(z)(A^L_MA^{L\ M}+A^R_MA^{R\ M})\right]\nn\\
&+\frac1{k_s}\int d^5x\sqrt{-g}e^{-\Phi(z)}\left[ \Tr (D_MH)^\dagger(D^MH)
-M^2_H(z) \Tr HH^\dagger - b(z) \Tr (H+H^\dagger)\right]. 
\eea

To deal with the mixing term between the axial vector fields and the pions we make a redefinition of the vector fields inspired by their would-be gauge transformation property (in order to keep $F_{MN}=\widehat F_{MN}$):
\bea
(A_L)_M= \xi_G^\dagger(\widehat A_L)_M \xi_G-i\partial_M\xi_G^\dagger \xi_G,\\ 
(A_R)_M= \xi_G(\widehat A_R)_M \xi_G^\dagger+i\xi_G \partial_M \xi_G^\dagger,\\
\xi_G=\exp\left(\frac{i\pi^a T^a}{\widehat\chi_\pi}\right).
\eea
From now on we call ``vector'' the fields $V^a=\frac{\widehat A_L^a+\widehat A_R^a}2$ 
and ``axial'' the orthogonal combination $A^a=\frac{\widehat A_R^a-\widehat A_L^a}2$. 
The parameter $\widehat\chi_\pi$ is tuned in order to eliminate the mixing:
\be
\widehat\chi_{\pi}=-\chi_\pi (1+\beta),\quad \beta=\frac{k_s}{4g_5^2}\frac{M^2(z)}{f^2(z)}.
\ee
We assume that the factor $\beta$ introduced here has no $z$-dependence. 
That is crucial to the determination of the possible $z$-dependencies of $f(z)$ and $b(z)$. The limit $\beta=\infty$ corresponds to the absence of the spontaneous breaking and signifies the restoration of the chiral symmetry.

\section{Holographic equations of motion}
\label{5DEOM}
In holography one gets from the equations of motion (EOM) two types of solutions \cite{Balasubramanian1999,Klebanov_1999}.
Let us briefly describe their interpretation within the AdS/QCD framework for the case of a general field $\varphi(=s,\pi,V,A)$ and suppressing the Lorentz and group indices. 

The first type of solution is the bulk-to-boundary propagator and describes the evolution of the $5D$ field from its boundary value(=source): $\varphi(x,z)=\widehat\varphi(x,z) \phi_{\mathcal O}(x)$. To simplify the notation we further use the same symbol (no hat) for a $5D$ field and its propagator. The choice of solution is governed by the holographic prescription for its UV ($z=\varepsilon$) asymptotics and by the demand of no faster than the power-law growth in the IR ($z=\infty$). The latter is specific to the SW-type models.

Another type is the Kaluza-Klein (KK) solution. One can realize the $5D$ field in terms of the physical $4D$ degrees of freedom with proper quantum numbers, $\varphi_{(n)}(x)$, as $\varphi(x, z)=\sum\limits_{n=0}^\infty \varphi_n(z)\varphi_{(n)}(x)$, where the sum goes over the possible radial excitations $n$. The EOMs define the $z$-profiles, $\varphi_n(z)$, in these KK expansions.
These profiles are subject to a certain orthogonality condition that leads towards a canonically normalized kinetic term for the four-dimensional fields after the $z$ coordinate is integrated over in the holographic action. 


\subsection{Vector and axial vector fields}

Duality  establishes the field-operator correspondence and the UV behavior of the bulk-to-boundary propagators
\bea
V^a(x,\varepsilon)=1\cdot \phi_V^{a\ \mu}(x) \ &\leftrightarrow&\  g_V \mathcal O_{V\ \mu}^a(x)=g_V{\overline\Psi\gamma^\mu T^a\Psi},\\
A^a(x,\varepsilon)=1\cdot \phi_A^{a\ \mu}(x) \ &\leftrightarrow&\  g_V \mathcal O_{A\ \mu}^a(x)=g_V{\overline\Psi\gamma^\mu\gamma_5 T^a\Psi}.
\eea

We work in a holographic gauge $A_z=V_z=0$ and $\partial_\mu A^\mu=\partial_\mu V^\mu=0$. The latter condition can be preserved on-shell only, and for the axial field it is necessary to have no mixing with the pions left.
The EOMs for the transverse part of the vector and axial vector fields are
\bea
\left(\partial_z \frac{e^{-\Phi}}z \partial_z V^{a}_\mu(x,z)-\frac{e^{-\Phi}}z \Box V_\mu^{a}(x,z)-\frac{M^2(z)R^2e^{-\Phi}}{z^3}V_\mu^{a}(x,z)\right)_\bot=0,\\
\left(\partial_z \frac{e^{-\Phi}}z\partial_z A_\mu^a(x,z)-\frac{e^{-\Phi}}z \Box A_\mu^a(x,z)-\frac{M^2(z)R^2e^{-\Phi}}{z^3}\frac{1+\beta}{\beta}A_\mu^a(x,z)\right)_\bot=0.
\eea

Analytic solutions can be achieved for an ansatz of the form
\be
M^2(z)R^2=4\mu_V\cdot \kappa^2z^2.\ee
The absence of the constant term is due to the holographic prescription for the vector mass in the UV, and it is a necessary choice for the correct behavior of the vector bulk-to-boundary propagator on the boundary. After the Fourier transformation, we obtain
 \be\label{D_btbprop1}
 V(q, z)=\Gamma\left(1-\frac{q^2}{4\kappa^2}+\mu_V\right)\Psi\left(-\frac{q^2}{4\kappa^2}+\mu_V,0;\kappa^2z^2\right),\ V(q,0)=1.
 \ee
 The special function $\Psi$, named after Tricomi, is the solution of the confluent hypergeometric equation with a proper behavior at $z$-infinity.
  The difference in the axial vector case consists just in a constant shift $\mu_V \rightarrow \mu_V \frac{1+\beta}{\beta}$; the axial vector propagator is:
\be
A(q, z)=\Gamma\left(1-\frac{q^2}{4\kappa^2}+\mu_V \frac{1+\beta}{\beta}\right)\Psi\left(-\frac{q^2}{4\kappa^2}+\mu_V \frac{1+\beta}{\beta},0;\kappa^2z^2\right)
,\ A(q,0)=1. \ee

 The parameter $\mu_V$ remains free and also appears in the expression of the normalizable solutions. The orthogonality relation is 
$\frac R{g_5^2}\int\limits_0^\infty dz e^{-\kappa^2z^2}z^{-1}V_n/A_n(z)V_k/A_k(z)=\delta_{nk}$. Then the $z$ profiles are determined from the EOMs and the spectra can be expressed using the discrete parameter $n=0,1,2,...$:
\bea
V_n (z)=A_n (z)=\kappa^2z^2\sqrt\frac{g_5^2}R \sqrt\frac{2}{n+1} L_n^1(\kappa^2z^2),\\
M_V^2(n)=4\kappa^2(n+1+\mu_V),\quad
M_A^2(n)=4\kappa^2\left(n+1+\mu_V +\frac{\mu_V}{\beta}\right). \label{Vspectrum}
\eea
Here $L_n^m(\kappa^2z^2)$ are the generalized Laguerre polynomials.  These solutions are analogous to those obtained in the standard framework after $\mu_V\rightarrow0,\ \frac{\mu_V}{\beta}\rightarrow$ constant. Linearity of the radial Regge trajectories $M^2(n)\sim n$ is a distinctive feature of the SW model and indicates a proper realization of confinement.

The quantum numbers of the corresponding operators allow us to identify the boundary fields $(V/A)_{(n)}(x)$ and the masses $M_{V/A}(n)$ with the massive radial excitations of $\rho$ and $a_1$ mesons. 

Let us consider also an alternative treatment. Having computed the Green's function $G(q,z,z')=\sum\limits_n\frac{\varphi_n^*(z)\varphi_n(z')}{q^2-M^2(n)}$, one can arrive at the following expression for the propagators
\bea\label{D_btbprop2}
V(q,z)=\sum\limits_n\frac{F_V(n)V_n(z)}{-q^2+M^2_V(n)},
\
A(q,z)=\sum\limits_n\frac{F_A(n) A_n(z)}{-q^2+M^2_A(n)},\\ 
F_A^2(n)=F_V^2(n)=\frac{8R\kappa^4}{g_5^2}(n+1).
\eea
It can be proved that the UV boundary conditions are respected in this form as well.

Therefore, we have determined two kinds of phenomenologically relevant quantities: the masses and the decay constants related to the states in the vector and axial vector sectors. 
The following matrix elements define the experimentally observed  quantities $F_\rho$ and $F_{a_1}$:
\bea\label{ZofD}
\langle0|\mathcal O_V^{a\ \mu}(x)|\rho^b(p)\rangle&=&\epsilon^\mu \delta^{ab} F_\rho e^{-ipx}\equiv\epsilon^\mu \delta^{ab} \frac1{g_V} F_{D}(0) e^{-ipx},\\
\langle0|\mathcal  O_A^{a\ \mu}(x)|a_{1}^b(p)\rangle&=&\epsilon^\mu \delta^{ab} F_{a_1} e^{-ipx}\equiv\epsilon^\mu \delta^{ab} \frac1{g_V} F_{D}(0) e^{-ipx}.
\eea
In our model, though the masses in the vector and axial vector channels are different, their decay constants coincide, while experimentally they are known to be distinct.
The experimental value of $F_\rho$ is estimated from the $\rho\rightarrow e^+e^-$ decay rate~\cite{PDG2018}, and $F_{a_1}$  could be obtained from the study of the $\tau$ decays \cite{PhysRevD.39.1357}.

\subsection{Scalar and pseudoscalar fields}
Let us follow similar steps in the case of spin zero fields. Due to the specifics of the linearized form of the $H$ field
\be\label{LinH}
H(x,z)=f(z)+s^a(x,z)T^a+\frac{2if(z)}{\chi_\pi}\pi^a(x,z)T^a,
\ee
 the correspondence in the scalar sector is the following
\bea \label{ScalarAsympt}
s^a(x,\varepsilon)=\frac\varepsilon{R} \phi^a_{S}(x)
\ &\leftrightarrow& \ g_S\mathcal O_S^a(x)= g_S\overline\Psi T^a\Psi,\\ \label{PScalarAsympt}
\pi^a(x,\varepsilon)=-\frac\varepsilon{R} \frac{\chi_\pi}{2f(\varepsilon)}\phi^a_{P}(x)\ &\leftrightarrow&\ g_S\mathcal O_P^a(x)= g_S \overline\Psi i\gamma_5T^a\Psi.
\eea
The associated QCD states are $a_0$ and $\pi$ mesons.

The EOMs for the scalar and pseudoscalar fields are
\bea
\partial_z \frac{e^{-\Phi}}{z^3}\partial_zs^a -\frac{e^{-\Phi}}{z^3} \Box s^a
-\frac{M^2_H(z)R^2}{z^5}e^{-\Phi}s^a=0,\\
\partial_z \frac{e^{-\Phi}}{z^3}{f^2(z)}\partial_z\pi^a-\frac{e^{-\Phi}}{z^3}f^2(z) \Box\pi^a+\frac{b(z)f(z)R^2}{z^5}e^{-\Phi}\frac{1+\beta}{\beta}\pi^a=0.
\eea

In the pseudoscalar case we have to choose a function $b(z)$. The function $f(z)$ is already uniquely fixed by the ansatz selected for $M^2(z)$,
\be\label{fAnsatz}
f(z)R=\sqrt{\frac{k_s}{g_5^2}\frac{\mu_V}{\beta}} \cdot \kappa z.
\ee

The condition~(\ref{fAnsatz}) allows us to write the pion EOM in a form reminiscent of the vector EOM
\be
\partial_z \frac{e^{-\Phi}}{z}\partial_z\pi^a-\frac{e^{-\Phi}}{z} \Box\pi^a+\frac{e^{-\Phi}}{z^3}(b_1+4b_2\cdot \kappa^2 z^2)\pi^a=0,
\ee
where we have assumed that the function $b(z)$ is chosen so that
\be\label{bAnsatz}
b(z)R^3\cdot (1+\beta)\sqrt\frac{g_5^2}{ k_s \mu_V\beta}=b_1 \kappa z +4b_2\cdot \kappa^3 z^3.
\ee
Any higher order terms would result in a non-analytic solution. We must impose 
$b_1=0$ in order to fulfill the boundary condition of Eq.~(\ref{PScalarAsympt}).  
Then, the bulk-to-boundary propagators are
\bea\label{ScProp}
s(q,z)=\frac z{R}\Gamma\left(\frac32+\mu_H-\frac{q^2}{4\kappa^2}\right)\Psi\left(\frac12+\mu_H-\frac{q^2}{4\kappa^2},0;\kappa^2z^2\right),\\
\pi(q,z)=-\sqrt\frac{g_5^2\beta}{k_s \mu_V} \frac{\chi_\pi}{2\kappa}\Gamma\left(1-b_2-\frac{q^2}{4\kappa^2}\right)\Psi\left(-b_2-\frac{q^2}{4\kappa^2},0;\kappa^2z^2\right).
\eea

The EOMs and the orthogonality conditions,
\bea
\frac {R^3}{k_s} \int\limits_0^\infty dz e^{-\kappa^2z^2} z^{-3}s_n(z)s_k(z)=\delta_{nk},\\
\frac{4\beta}{(1+\beta)\chi_\pi^2}\frac{R^3}{k_s}\int\limits_0^\infty dz e^{-\kappa^2z^2}z^{-3}f^2(z)\pi_{n}(z)\pi_{k}(z)=\delta_{nk},
\eea
bring the following solutions for the KK $z$-profiles
\bea
s_n(z)=\frac zR \sqrt\frac{k_s}R \sqrt{\frac{2}{n+1}}(\kappa z)^2L^1_n(\kappa^2z^2),\quad M_s^2(n)=4\kappa^2(n+3/2+\mu_H),\\
\pi_n(z)=\frac{\chi_\pi}{\kappa}\sqrt\frac{1+\beta}{\mu_V} \sqrt\frac{g_5^2}{2R}(\kappa z)^2L^1_n(\kappa^2z^2),\quad  M_{\pi}^2(n)=4\kappa^2\left(n+1- b_2\right).
\eea
 Assuming $b_2=1$ makes the ground state Goldstones massless, $m_\pi=M_\pi(0)=0$. This reveals the goal of $b(z)$ introduced in the scalar potential of the $5D$ action: with the analyticity of the solution imposed, it only serves to nullify the pion masses. However, even without it, we can generally distinguish the $m_\rho=M_V(0)$ and $m_\pi$ scales due to the appearance of $\mu_V$ in the vector masses.
 Notice that we gain an analytic result for the whole tower of pion radial excitations, while in most holographic papers one finds an implicit equation defining numerically just the ground state.

The alternative expressions for the propagators are analogous to the ones found in the vector sector
\bea\label{ScProp2}
s(q,z)=\frac1{\sqrt2}\sum\limits_n\frac{F_s(n)s_n(z)}{-q^2+M^2_s(n)},
\ \pi(q,z)=\sum\limits_n\frac{F_\pi(n)\pi_n(z)}{q^2-M^2_\pi(n)},\\ \label{Fofs}
F_s^2(n)=16\kappa^4\frac{R}{k_s}(n+1),\ F_\pi^2(n)=8\kappa^4\frac\beta{1+\beta}\frac{R}{k_s}(n+1).
\eea
The factor $1/\sqrt{2}$ in front of the scalar propagator is necessary to conform to the usual definition of
the scalar decay constant. The true value of the decay constant is found only after one calculates the residue at
$q^2 = M^2(n)$ of the corresponding two-point function. We follow the conventions of
Ref.~\cite{Colangelo2008} and we use their definition of $F_s$. In Section \ref{2ptCorrelator}, we will re-encounter this quantity in the
residue of the scalar correlator, and the $1/\sqrt{2}$ factor ensures the agreement between both expressions.

The quantities in the last equations above are related to the decay constants $F_s$ and $F_\pi$ appearing in the one-point functions
\bea
\langle 0|\mathcal O_S^a(x)|a_0^b\rangle=\delta^{ab} F_s e^{-ipx}\equiv\delta^{ab}\frac1{g_S}F_s(0)e^{-ipx},\\
\langle 0|\mathcal O_P^a(x)| \pi^b\rangle=\delta^{ab} F_\pi e^{-ipx}\equiv\delta^{ab}\frac1{g_S}F_\pi(0)e^{-ipx}.
\eea
The numerical information on the value of $F_s$ can be found in the phenomenological studies of \cite{Gokalp2001}. $F_\pi$ appears in various relations of the chiral perturbation theory, and in the chiral limit it can be related to the pion decay constant $f_{\pi}$ and the quark condensate through the condition $f_{\pi} F_\pi=-\langle0|q\bar q|0\rangle$ \cite{GASSER198277,*Gasser1984}\footnote{This condition appears in the chiral limit as a consequence of the equation that one gets considering the divergence of the axial vector current, $f_\pi m^2_\pi=F_\pi (m_u+m_d)$, 
and the Gell-Mann--Oakes--Renner relation, $f^2_\pi m^2_\pi=-(m_u+m_d)\langle0|q\bar q|0\rangle$.}. 

The numerical predictions for the decay constants are provided in Section~\ref{Experiment}.

\subsection{Dynamics and interpretation of $f(z)$}
 In this analysis, we would like to stay within the chiral limit, where on the QCD side the breaking is generated dynamically by the chiral condensate $\langle q\bar q\rangle$.
In the holographic bulk we have a sigma-model type theory, where the function $f(z)$ describes the spontaneous symmetry breaking in a non-dynamical fashion. 

However, there is no clear holographic prescription on how the chiral symmetry breaking should be realized. In fact, the specifics of the realization define wholly different classes of models, e.g. in the framework with the IR cut-off one can choose between those of Refs.~\cite{HW_2005}, \cite{daRold_2005} or \cite{Hirn2005}. 
In a general AdS/QCD framework (that of \cite{HW_2005, SW_2006}) the conventional understanding is that the scalar vev has the following form (see also \cite{Erlich2008})
\be \label{canonic_f_vev}
f(z)R=m_qz+\frac\sigma 4 z^3,
\ee
where the parameters $m_q$ and $\sigma$ are believed to correspond to the physical current quark mass and the chiral condensate. 
This power behavior is a solution of the EOM written for $f(z)$ in the case of the HW model with $\Phi(z)=0$, 
while in the SW the powers get multiplied by the hypergeometric functions (see below).
The interpretation in Eq.~(\ref{canonic_f_vev}) is motivated by the AdS/CFT correspondence \cite{Balasubramanian1999,Klebanov_1999}: $m_q$ is the physical source for the $\mathcal O=q\bar q$ operator and $\sigma$ is a vev determined as a one-point function in the presence of a source, $\langle\mathcal O\rangle_\phi$. That means that if the source($=m_q$) goes to zero, the vev vanishes in the case of the normal-ordered observables $\langle\mathcal O\rangle_{\phi=0}=0$. One has to admit that this is not compatible with QCD where the chiral condensate is non-zero in the chiral limit. Most authors do not try to explain this issue, though in the HW setup of Ref.~\cite{DaRold_scalar} they introduce an extra scalar potential on the IR brane to get around the problem. 

In the SW the function form (\ref{canonic_f_vev}) is not a solution of the EOM, but it is a common opinion that it should emerge in the UV asymptotics at least. The problem  arises that while
 choosing a solution finite at $z\rightarrow\infty$, one is left with only one branch of the equation.  Hence, the model bears a correlation in the definition of the coefficients at $z$ and $z^3$ terms, mixing the coefficients associated in QCD with the explicit and spontaneous sources of the breaking. Various attempts were made to resolve this contradiction: from manually inserting a different ansatz \cite{Kwee2008} towards major modifications of the model dilaton and/or scalar potential to make a consistent dynamical solution for $f(z)$\cite{Gherghetta2009,Sui2010,Vega2010}. The latter models give independent predictions for $m_q$ and $\sigma$, but in our opinion, they are no longer compatible with the strict AdS/CFT identification, not to mention its unclear realization in the chiral case.  

It is evident that our ansatz for $f(z)$ given in Eq.~(\ref{fAnsatz}) does not follow the form of Eq.~(\ref{canonic_f_vev}). Nevertheless, the appearance of Eq.~(\ref{fAnsatz}) is related to the correct description of the vector sector. And we put reasonings on the analyticity and holographic consistency of the previous sections prior to the issue of possible identifications of the $f(z)$ parameters, especially in light of the discussion presented above. Let us mention several other arguments. First,  it could be reasonable to demand $f(z\rightarrow\infty)R\sim z$ (as is done in Ref.~\cite{Gherghetta2009}) that fixes the parallel slopes of the vector and axial vector trajectories
in accordance with the idea of the chiral symmetry not being restored \cite{Afonin2007, Shifman2008}. We may attribute our ansatz (\ref{fAnsatz}) to the preservation of this quality in a simple manner. Second, one can speculate that a mass appearing at the linear in $z$ order is not a current but a constituent one \cite{Afonin2011}, that light quarks acquire in the presence of the quark condensate. We will show that, indeed, the factor could be of an order $\sim300$~MeV for a natural value of $g_S$. And finally, we can refer to Ref.~\cite{Cherman2009}, in which it is concluded that because the scale dependence is not systematically dealt with in the bottom-up holographic models, it might be advisable to give up on matching to such quantities as $m_q$ and $\sigma$.

With a firm resolution to use the ansatz of Eq.~(\ref{fAnsatz}), let us nevertheless explore the case where $f(z)$ is a solution of the EOM. In our model this is not quite standard: there is a new coefficient $\mu_H$ and the scalar potential with $b(z)$ makes the equation inhomogeneous,
  \be \label{f_zEOM}
\partial_z \frac{e^{-\Phi(z)}}{z^3}\partial_zf(z)-\frac{e^{-\Phi(z)} M^2_H(z)R^2}{z^5}f(z)-\frac{b(z)}{z^5}e^{-\Phi(z)}=0.
\ee
The homogeneous part coincides with the EOM of a conventional SW but for an addition of $\mu_H$. The solution changes accordingly,
$$f_{hom}(z)\sim (\kappa z)^3 \cdot \ _1F_1\left(\frac32+\mu_H,2,\kappa^2 z^2\right) + \kappa z\cdot \Psi\left(\frac12 +\mu_H,0;\kappa^2 z^2\right),$$
where $\ _1F_1$ and $\Psi$ are confluent hypergeometric functions of different types.

With  $b(z)$ taken from Eq.~(\ref{bAnsatz}) (though we might have used any arbitrary coefficient function $\sim b_1 z+ b_2 z^3$, it would be necessary to have $b_1=0$ to get a finite result), the particular solution turns out to be (with the use of the relevant Green's function $G(z,z')$):
\bea
f_{part}(z)R&=&\int\limits_0^\infty dz'\frac{b(z')e^{-\Phi(z')}}{z'^5}G(z,z')\nn\\
&=&\frac{-\kappa b_2}{1+\beta}\frac{z}{\kappa^2R^2}\sqrt{\frac{k_s}{g_5^2}\mu_V\beta}\left[\frac1{\mu_H+1/2}+\Gamma\left(\mu_H+\frac12\right)\Psi\left(\mu_H+\frac12,0;\kappa^2 z^2\right)\right].\eea
We can see that  for $f(z)=f_{hom}(z)+f_{part}(z)$ a $f(z)R\sim z$ approximation is an appropriate one if we keep just the leading asympotics for $z\rightarrow0$. Additionally, we have a separate source for the $\sim z$ terms aside from those coming from the Tricomi function.

Moreover, for specific values of $\mu_H$ we can simplify the EOM~(\ref{f_zEOM}) so that a solution of the homogeneous part,
that is finite in the IR, is either linear ($\sim z$) at $\mu_H=-1/2$ or cubic ($\sim z^3$)  at $\mu_H=-3/2$.
The case $\mu_H=-1/2$ seems most interesting, as it would prove our choice of the ansatz if no $b(z)$ were present; though the full solution is 
$
f(z)\sim C_{hom} z+C_{part} z \ln z.
$
Furthermore, $\mu_H=-1/2$ makes the scalar tower $M_s^2(n)=4\kappa^2(n+1)$ look exactly like a shifted pseudo-scalar one, meaning $m_{a_0}=m_{\pi'}$. A finite pion mass could be a source of the splitting between them. We will use the assumption of fixing $\mu_H=-1/2$ in one of the phenomenological fits.

\section{Two-point correlators}\label{2ptCorrelator}

Following the duality connection between the $4D$ partition function and the on-shell holographic action we present a definition for the two-point functions, with $\mathcal O_\mu$ standing for spin one operators and $\mathcal O$ for spin zero,
\bea
&\langle g_V \mathcal O_\mu^{a}(q) g_V \mathcal O_\nu^{b}(p)\rangle=\delta(p+q)\int d^4x e^{iqx}\langle g_V \mathcal O_\mu^{a}(x) g_V \mathcal O_\nu^{b}(0)\rangle
=\frac{\delta^2 iS_{5D}^{on-shell}}{\delta i \phi^{a}_{\mu}(q)\delta i \phi^{b}_{\nu}(p)},\\
&i\int d^4x e^{iqx}\langle g_V \mathcal O_\mu^{a}(x) g_V \mathcal O_\nu^{b}(0)\rangle=\delta^{ab}\left(\frac{q_\mu q_\nu}{q^2}-\eta_{\mu\nu}\right)\Pi_{V,A}(q^2),\label{V_cor_def}\\
&i\int d^4x e^{iqx}\langle g_S \mathcal O^a(x) g_S \mathcal O^b(0)\rangle=\delta^{ab}\Pi_{s,\pi}(q^2).\label{Sc_cor_def}
\eea

It is known that there could be divergences present in the functions of this type.
If we perform a simple short-distance $\varepsilon$ cut-off regularization as $z\rightarrow0$ the resulting expressions are the following:
\bea\label{V2pt_epsilon}
&\Pi_{V}(q^2)=\frac{2\kappa^2R}{g_5^2} \left(\mu_V-\frac{q^2}{4\kappa^2}\right) \left[\ln \kappa^2\varepsilon^2+2\gamma_E+\psi\left(1+\mu_V-\frac{q^2}{4\kappa^2}\right)\right],\\
&\Pi_{A}(q^2)=\frac{2\kappa^2R}{g_5^2}\left(\mu_V+\frac{\mu_V}{\beta}-\frac{q^2}{4\kappa^2}\right)\left[\ln \kappa^2\varepsilon^2+2\gamma_E+\psi\left(1
+\mu_V\frac{1+\beta}{\beta}-\frac{q^2}{4\kappa^2}\right)\right],\\\label{S2pt_epsilon}
&\Pi_{s}(q^2)=\frac{4\kappa^2R}{k_s}\left(\frac12+\mu_H-\frac{q^2}{4\kappa^2}\right) 
\left[\ln \kappa^2\varepsilon^2+2\gamma_E-\frac12+\psi\left(\frac32+\mu_H-\frac{q^2}{4\kappa^2}\right)\right],\\
\label{PS2pt_epsilon}
&\Pi_{\pi}(q^2)=\frac{2\kappa^2R}{k_s}\frac\beta{1+\beta}\left(-1-\frac{q^2}{4\kappa^2}\right) \left[\ln \kappa^2\varepsilon^2+2\gamma_E+\psi\left(-\frac{q^2}{4\kappa^2}\right)\right].
\eea
The $\Pi_{s}$ correlator also possesses a $\varepsilon^{-2}$ singularity that is eliminated after the proper counterterm at the boundary is introduced.
With the series representation of the digamma function $\psi\left(\frac32+\mu_H-\frac{q^2}{4\kappa^2}\right)=-\gamma_E+\sum\frac1{n+1}+\sum \frac{4\kappa^2}{q^2-M^2_s(n)}$ we can check that the residue of $\Pi_{s}$ is a quantity equal to $F_s^2$ as defined in Eq.~(\ref{Fofs}). The same procedure validates other decay constants.

Alternatively (and in need of a regularization) we can express the correlators as
\bea \label{resonanceV2pt}
\Pi_{V}(q^2)=\sum\limits_{n=0}^\infty\frac{F^2_V(n)}{-q^2+M^2_V(n)},\ \Pi_{A}(q^2)=\sum\limits_{n=0}^\infty\frac{F^2_A(n)}{-q^2+M^2_A(n)},\\
\label{resonanceS2pt}
\Pi_{s}(q^2)=\sum\limits_{n=0}^\infty\frac{F^2_s(n)}{-q^2+M^2_s(n)},\ \Pi_{\pi}(q^2)=\sum\limits_{n=0}^\infty\frac{F^2_\pi(n)}{-q^2+M^2_\pi(n)}.
\eea
These expressions can be achieved using Eqs.~(\ref{D_btbprop2}) and (\ref{ScProp2}). Though in the case of $\Pi_{s}$ the explicit derivation with $s(q,z)$ of Eq.~(\ref{ScProp2}) leads as well to the $\varepsilon^{-2}$ singularity and a non-pole term\footnote{The derivation of the scalar two-point function, both in Eq.~(\ref{S2pt_epsilon}) and in Eq.~(\ref{resonanceS2pt}), stands out among other cases. In $\Pi_{s}(q^2)\sim \varepsilon^{-3}s(q,\varepsilon)\partial_z s(q,\varepsilon)$, one has to include several orders in the series: $s(q,\varepsilon)\sim \varepsilon + \varepsilon^3$ and $\partial_z s(q,\varepsilon)\sim \varepsilon^0 + \varepsilon^2$. The estimation of $\partial_z s(q,\varepsilon)$ should be performed carefully in the case of the definition of Eq.~(\ref{ScProp2}) because of taking the small$-z$ limit inside the infinite sum.}; both are suppressed in Eq.~(\ref{resonanceS2pt}).

It is evident that the correlators of Eqs.~(\ref{V2pt_epsilon} -- \ref{PS2pt_epsilon}) and those of Eqs.~(\ref{resonanceV2pt}, \ref{resonanceS2pt}) differ. However, it could be shown that the differences are encoded within the polynomial structure of a type $C_0 +C_1 q^2$. These are the known ambiguities of a two-point function. With those subtracted, we arrive at the convergent correlator that has a similar structure in all the cases,
\be
\widehat\Pi(q^2)=\sum\limits_{n=0}^\infty\frac{q^4 F^2(n)}{M^4(n)(-q^2+M^2(n))}.
\ee

The most interesting and assumed regularization independent quantity in the spin one sector is the left-right combination $\Pi_{LR}$:
\be
\Pi_{LR}(q^2)= \Pi_{V}(q^2)-\Pi_{A}(q^2).
\ee

In the region of  small Euclidean momenta ($Q^2=-q^2$) at the
$(Q^2)^0$ order we obtain from $\Pi_{LR}$ a constant coefficient that we call $F^2$. Both vector and axial vector correlators have some non-zero constant factor at this order. Their difference should establish the one free of the short-distance ambiguities. Nevertheless, the final quantity still contains the $\varepsilon$ divergence:
\be
F^2=\frac{2R \kappa^2 \mu_V}{g_5^2}\left[\psi\left(1+\mu_V\right)-\psi\left(1+\mu_V\frac{1+\beta}{\beta}\right)-\frac1\beta\left(\ln\kappa^2\varepsilon^2+2\gamma_E+\psi\left(1+\mu_V\frac{1+\beta}{\beta}\right)\right) \right]
\ee
Otherwise it can be represented as a divergent series
\be\label{pionDecay}
F^2=\sum\limits_n\frac{F^2_V(n)\cdot 4\kappa^2\mu_V/\beta}{M^2_V(n)M^2_A(n)}
=\frac{2R \kappa^2 \mu_V}{g_5^2\beta}\sum\limits_n\frac{n+1}{(n+1+\mu_V)(n+1+\mu_V+\mu_V/\beta)}.\ee

In QCD one finds a definition of $f_{\pi}$, the pion decay constant in the chiral limit, in the matrix element
\be\label{f_pi_def}
\langle0|\bar q \gamma_\mu\gamma_5 T^a q(0)|\pi^b(p)\rangle=ip_\mu f_{\pi} \delta^{ab}
\ee
The experimental value is $f_{\pi}=92.07\pm1.2$~MeV~\cite{PDG2018}. To make the connection to the model-defined coefficient $F$, we have to first introduce some regularization in the latter, and second, take into account that the operators used in the construction of $\Pi_{LR}$  differ from the one of Eq.~(\ref{f_pi_def}) by the yet undetermined factor $g_V$\footnote{The factors $g_V$ and $g_S$ appear due to the conventions taken in Eqs.~(\ref{V_cor_def},\ref{Sc_cor_def}). It turns out that they are reabsorbed (using the matching conditions of Eq.~(\ref{g_VS_matching})) in the physical parameters of this Section, but not in those related to the three-point correlators.}.
Let us assume a vector meson dominance (VMD) like regularization, meaning cutting the sum in Eq.~(\ref{pionDecay}) at the first term. Further we will use this VMD limit to estimate the experimental observable as $f_{\pi}=F_{reg}/g_V $. We will see that this assumption brings a good result for $f_{\pi}$.

The next term in the small-$Q^2$ expansion, $(Q^2)^1$ order, brings the $L_{10}$ coefficient.
 \begin{align} \label{L10}
& g^2_V L_{10}=\frac14 \left.\frac d{dQ^2}\left(\Pi_V(Q^2)-\Pi_A(Q^2)\right)\right|_{Q^2=0}\\ \nn
 &=\frac{R }{8g_5^2} \bigg[\psi\left(1+\mu_V\right)-\psi\left(1+\mu_V\frac{1+\beta}{\beta}\right)
 +\mu_V\psi_1\left(1+\mu_V\right)-\mu_V\frac{1+\beta}{\beta} \psi_1\left(1+\mu_V\frac{1+\beta}{\beta}\right)\bigg].
  \end{align}
  The phenomenological value of $L_{10}$ at the scale of the $\rho$ mass is $(-5.5\pm0.7)\cdot 10^{-3}$ \cite{Pich1995}.
  
Now let us consider the high-energy asymptotics of the calculated two-point functions. The QCD result stemming from the operator product expansion (OPE) is well-known \cite{SHIFMAN1979385, REINDERS1985}:
\begin{align}\label{QCD2pt}
\Pi_{V,A}(Q^2)/Q^2=\frac{N_c}{24\pi^2}\left(1+\frac{\alpha_s}\pi\right)\ln\frac{Q^2}{\mu^2}-
\frac{\alpha_s}{24\pi}\frac{N_c}3\frac{\langle G^2\rangle}{Q^4}+c_{V,A}\frac{14N_c}{27}  \frac{\pi\alpha_s\langle q\bar q\rangle^2}{Q^6},\\\label{QCD2ptS}
\Pi_{s,\pi}(Q^2)/Q^2=\frac{N_c}{16\pi^2}\left(1+\frac{11\alpha_s}{3\pi}\right)\ln\frac{Q^2}{\mu^2}+
\frac{\alpha_s}{16\pi}\frac{N_c}3\frac{\langle G^2\rangle}{Q^4}-c_{s,\pi}\frac{11N_c}9  \frac{\pi\alpha_s\langle q\bar q\rangle^2}{Q^6},\\
c_V=1,\ c_A=-\frac{11}7,\ c_s=1,\ c_\pi=-\frac7{11}. \nn
\end{align}
These are computed for the operators with $g_V=g_S=1$. 
 $\langle G^2\rangle$ and $\langle q\bar q\rangle$ are the gluon and quark condensate, and $\alpha_s$ is the strong coupling constant. 
The scale dependent quantities here are usually estimated at the scale of chiral symmetry breaking  $\sim 1$~GeV: $\langle\frac{\alpha_s}\pi G^2\rangle=0.012$~GeV$^4$~\cite{SHIFMAN1979385} (lattice: $\langle\frac{\alpha_s}\pi G^2\rangle=0.10$~GeV$^4$ \cite{CAMPOSTRINI1989393}) and $\langle q\bar q\rangle=-(242\pm15)^3$~MeV$^3$~\cite{Jamin2002}  or $-(235\pm15)^3$~MeV$^3$~\cite{Colangelo2001}.

The results from our model are the following (assuming that the logarithm regularization in Eq.~(\ref{V2pt_epsilon}), in fact, can only be made up to a subtraction constant $\ln(Q^2\varepsilon^2)\rightarrow \ln\frac{Q^2}{\mu^2}+\lambda$)
\begin{align}\label{V2pt}
\Pi_{V}(Q^2)/Q^2=\frac{R}{2g_5^2} &\left\{\ln\frac{Q^2}{\mu^2}+\lambda_V+\frac{2\kappa^2}{Q^2}\left[1+2 \mu_V \left(\ln\frac{Q^2}{\mu^2}+\lambda_V+1\right)\right] \right.\\
	&\left.+\frac{4\kappa^4}{3Q^4}\left[-1+6\mu_V^2\right]+\frac{16\kappa^6}{3Q^6}\mu_V\left[1-2\mu_V^2\right]+\mathcal O\left(\frac1{Q^{8}}\right)\right\},\nn
\end{align}
and $\Pi_{A}(Q^2)$ is given by a similar expression with the change $\mu_V \rightarrow \mu_V+\frac{\mu_V}{\beta}$. For the spin zero two-point functions we have
\bea\label{S2pt}
\Pi_{s}(Q^2)/Q^2&=&\frac{R}{k_s} \left\{\ln\frac{Q^2}{\mu^2}+\lambda_S+\frac{2\kappa^2}{Q^2}\left[1 +\left(1+2\mu_H\right) \left(\ln\frac{Q^2}{\mu^2}+\lambda_S+1\right)\right] \right.\\
	&&\left.\quad+\frac{2\kappa^4}{3Q^4}\left[1+12\mu_H(1+\mu_H)\right]+\frac{4\kappa^6}{3Q^6}\left[1+2\mu_H\right]\left[1-4\mu_H(1+\mu_H)\right]+\mathcal O\left(\frac1{Q^{8}}\right)\right\},\nn\\ \label{PS2pt}
\Pi_{\pi}(Q^2)/Q^2&=&\frac{R}{2k_s} \frac\beta{1+\beta} \left\{\ln\frac{Q^2}{\mu^2}+\lambda_P+\frac{4\kappa^2}{Q^2}\left[\ln\frac{Q^2}{\mu^2}+\lambda_P+\frac12\right]+\frac{20\kappa^4}{3Q^4}\right.\\
&&\left.\qquad\qquad\quad+\frac{16\kappa^6}{3Q^6}+\mathcal O\left(\frac1{Q^{8}}\right)\right\}.\nn
\eea

Matching the corresponding leading logarithmic terms in Eqs.~(\ref{V2pt},\ref{S2pt}) and Eqs.~(\ref{QCD2pt},\ref{QCD2ptS}) provides the values of the $5D$ coupling constants
\be\label{g_VS_matching}
g_V^2\frac{g_5^2}{R}=\frac{12\pi^2}{N_{c}},\quad g_S^2\frac{k_s}{R}=\frac{16\pi^2}{N_{c}}.
\ee
However, the scalar and pseudoscalar correlators have different asympotics, and an alternative expression from matching  Eqs. (\ref{QCD2ptS}) and (\ref{PS2pt}) could be $g_S^2\frac{k_s}{R}=\frac\beta{1+\beta} \frac{8\pi^2}{N_{c}}$. The results for $k_s$ coincide for $\beta=-2$ or in the case of the chiral restoration at $\beta=\infty$. Thus, we reach the conclusion that the consistency of the large-$Q^2$ asymptotics in the scalar sector fixes one of the model parameters to $\beta=-2$. We will see further that even in a global fit to the physical observables where $\beta$ is allowed to vary its value settles close to this one.

For the left-right correlator, the model gives
\begin{align}\label{LR_large_exp}
\Pi_{LR}(Q^2)/Q^2=-\frac{2R}{g_5^2}&\Bigg\{\frac{\kappa^2}{Q^2} \frac{\mu_V}\beta \left(\ln\frac{Q^2}{\mu^2}+\lambda_V+1\right)\\
&+\frac{2\kappa^4}{Q^4}\mu_V^2\frac{1+2\beta}{\beta^2} +\frac{4\kappa^6}{3Q^6} \frac{\mu_V}{\beta}\left[1-2\mu_V^2\left(3+\frac3\beta+\frac1{\beta^2}\right)\right]
+\mathcal O\left(\frac1{Q^{8}}\right)\Bigg\},\notag
\end{align}
Following Eq.~(\ref{QCD2pt}) we are supposed to obtain the manifestation of chiral symmetry breaking  $-\eta_{\mu\nu}\Pi_{LR}(Q^2)/g_V^2=-\eta_{\mu\nu}\frac{4\pi \alpha_s}{Q^4} \langle q \bar q\rangle^2$, while the other terms should vanish in the chiral limit. The relevant combination is estimated in Ref.~\cite{NARISON2005223}: $-4\pi\alpha_s\langle q\bar q\rangle^2=-(1.0\pm0.2)\times 10^{-3}$~GeV$^6$ (in the chiral limit), and 
in the holographic model we have
\be\label{V_ch_cond}
-4\pi\alpha_s \langle q\bar q\rangle^2 =\frac{8\kappa^6}{3}\frac{R}{g^2_V g_5^2}  \frac{\mu_V}{\beta}\left[1-2\mu_V^2\left(3+\frac3\beta+\frac1{\beta^2}\right)\right].
\ee
The other terms in Eq.~(\ref{LR_large_exp}) have no counterpart in the chiral limit of QCD: $\lambda$ in the  logarithm regularization  can be tuned to provide any constant piece in the $1/Q^2$ term, but the origin of $\ln Q^2/Q^2$ cannot be explained (the problem also encountered in \cite{GSW}); and the $1/Q^4$ term can only be related to $m_q \langle q\bar q\rangle$.

It is a common problem that the holographic models fail to be a match to QCD in these large-$Q^2$ expansions of the correlators even on a qualitative level. In the setups with an IR cut-off \cite{HW_2005, daRold_2005,Hirn2005} one faces the absolute lack of the next-to-leading order terms in the expansion, and the provided explanation is that the vector sector does not feel the symmetry breaking  due to the scalar vev and the breaking effect of the cut-off is decoupling exponentially fast at high energies. Later it was proposed to introduce the condensates by hand in Ref.~\cite{Hirn2006} or through a dynamical scalar with appropriate mass terms and potential coupled to gravity in a braneless approach in Ref.~\cite{Csaki2007}. 

In the conventional SW model there appears no $1/Q^6$ term in the vector correlator. It is a general feature for the vector two-point functions saturated by the narrow resonances with a spectrum of a type $\sim\kappa^2(n+1)$ \cite{Afonin2003, Afonin2004,Afonin2006}. The left-right correlator in the SW acquires an order parameter of the chiral symmetry breaking only from the axial vector contribution. There are several propositions to make an improvement in the vector correlator \cite{GSW,Jugeau2013}, and the appearance of $\mu_V$ in the intercept of the spectrum (\ref{Vspectrum}) can be considered as a possible solution too. 

We can speculate on connecting separately the $1/Q^4$ and $1/Q^6$ terms in $\Pi_V$ and $\Pi_A$ to the  condensates, but that does not sound reasonable. For instance, the gluon condensate prediction is distinct in the two channels, in contradiction to Eq.~(\ref{QCD2pt}); nor do we find a constant ratio between the $1/Q^6$ terms. After all, the condensates should manifest themselves as a result of the conformality violation, and both the HW and SW models propose just the simplest ways of doing it -- maybe the leading order logarithm is the only term where enough precision can be claimed. 

The situation does not become more consistent in the case of spin zero two-point functions. The $1/Q^4$ term in Eq.~(\ref{QCD2ptS}), associated with the gluon condensate, is coincident in Eqs.~(\ref{S2pt}) and (\ref{PS2pt}) just in the case of $\mu_H=-3/2$, rendering the $a_0$ state massless. And the constant ratio between the terms at $1/Q^6$ power can only be achieved with a positive value of $\mu_H$, which is not in the least favored in other observables.

Let us instead consider an alternative large-$Q^2$ expansion using the two-point functions of Eqs.~(\ref{resonanceV2pt}) and (\ref{resonanceS2pt}). As was mentioned they are in need of the regularization, and we assume to make it by cutting the tower of resonances at some finite number $N_m$.
As the structure of the correlators~(\ref{resonanceV2pt},\ref{resonanceS2pt}) is the same, the following asymptotics is true for each one of them
\be\label{resonanceLargeQ2}
\lim\limits_{Q^2\rightarrow\infty}\Pi(Q^2)/Q^2=\frac{\sum\limits_{n=0}^{N_m}F^2(n)}{Q^4}-\frac{\sum\limits_{n=0}^{N_m}F^2(n)M^2(n)}{Q^6}+\mathcal O\left(\frac1{Q^8}\right).
\ee
This expression seems  more appealing than those in Eqs.~(\ref{V2pt}-\ref{PS2pt}): it has a unified form and there are no unexplicable terms. Furthermore, as in our model $F_V(n)=F_A(n)$ the large-$Q^2$ limit of $\Pi_{LR}/Q^2$ starts with $1/Q^6$. This is  translated to the spin zero case, where $F^2_s(n)=2\frac{1+\beta}\beta F^2_\pi(n)$ and the equality can be achieved for $\beta=-2$. This value of the $\beta$ factor we have already seen in the comparison of the leading logarithmic terms. However, these logarithmic asymptotics themselves do not appear in this type of regularization, they need to have the whole infinite tower. Another drawback is that the gluon condensate comes with the wrong sign in the spin one cases and the quark condensate -- in the vector and pseudoscalar channels.

Though these discrepancies are present,
the situation for the $1/Q^6$ term with this regularization turns out to be more phenomenologically relevant.  The coefficients at $1/Q^6$ power are the following:
\bea\label{resV_ch_cond}
\Pi_{LR}:&\ \frac{4\kappa^6\mu_V (N_m+1)(N_m+2)}{\pi^2\beta},\\
\label{resS_ch_cond}
\Pi_{s}:&\ -\frac{\kappa^6(N_m+1)(N_m+2)(9+4N_m+6\mu_H)}{\pi^2},\\ \label{resP_ch_cond}
\Pi_{\pi}:&\ -\frac{2\kappa^6 N_m(N_m+1)(N_m+2)}{\pi^2}\frac\beta{1+\beta}.
\eea
Note that in the VMD limit of $N_m=0$ there is no contribution of this order in the pion correlator due to $m_\pi=0$. However, the logarithmic-independent quantity of $\Pi_{LR}$ is not only correctly assessed in the qualitative behavior of its $\frac1{Q^2}$ expansion, but the estimate (\ref{resV_ch_cond}) in the VMD limit has a better agreement with  Ref.~\cite{NARISON2005223} than that of Eq.~(\ref{V_ch_cond}), as we will see in Section~\ref{Experiment}.

\section{Three-point couplings, pion and axial form factors}\label{3ptCorrelator}
The $\rho_n \pi_{n_1}\pi_{n_2}$ coupling is obtained from the $5D$ Lagrangian as an integral over the three KK $z$-profiles 
\be
g_{\rho_n, \pi_{n_1},\pi_{n_2}}=\frac R{k_s}\int dze^{-\Phi(\kappa^2z^2)}\frac{1}{z^3}\frac{2f^2(z)R^2\cdot\beta(1+2\beta)}{\chi_\pi^2(1+\beta)^2} V_n(z) \pi_{n_1}(z)\pi_{n_2}(z).
\ee
The calculation is straightforward for any given set of the radial numbers $n,n_1,n_2$. In the case that we are only interested in the ground state pions $n_1=n_2=0$, the result is
\be
g_{\rho_n, \pi,\pi}=\sqrt\frac{2g_5^2}{R(n+1)} \frac{1+2\beta}{1+\beta}(\delta_{n,0}-\delta_{n,1}).
\ee

\begin{figure}
  \includegraphics[scale=0.9]{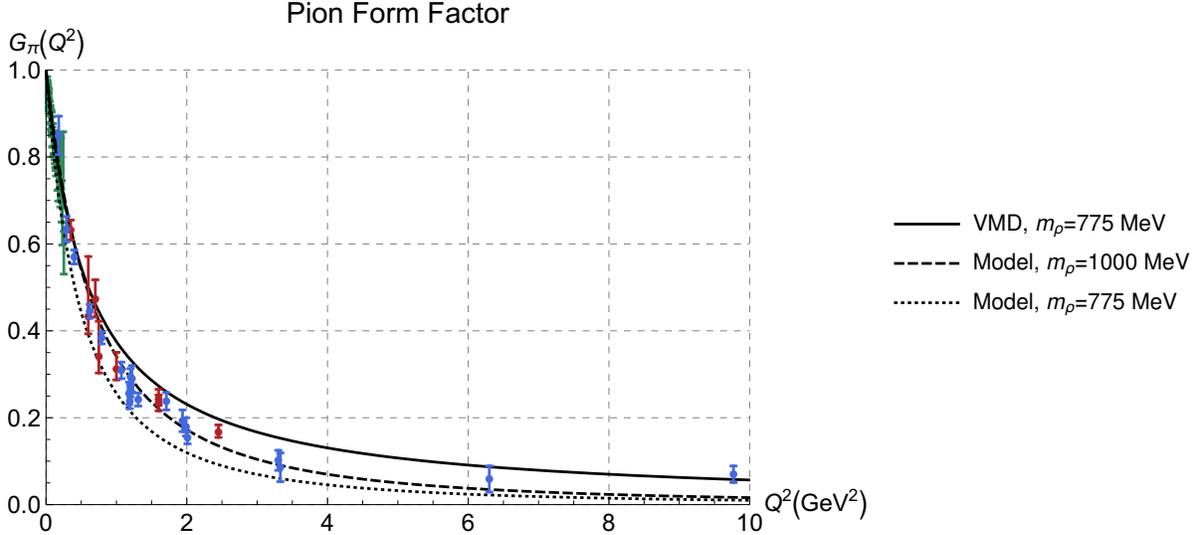}
    \caption{The pion form factor plot. The  experimental points belong to CERN \cite{AMENDOLIA1986168} (green), DESY and Jefferson Lab data \cite{Huber2008} (red), 
    and CEA/Cornell \cite{Bebek1978} (blue). The predicted lines are given for the cases  with one vector meson exchange (solid), and with two (the other two). The latter is the case of the model under consideration. $m_\rho=775$~MeV or $1000$~MeV and $m_{\rho'}=1465$~MeV were assumed.}
  \label{PionFFGraph}
\end{figure}

We also can examine the electromagnetic form factor (FF) of the pion $ G_\pi(q^2)$ defined as
\be
\langle \pi^a(k_1)| \mathcal O^{c\ \mu}_V (0)| \pi^b(k_2)\rangle = i\varepsilon^{abc}(k_1+k_2)^\mu G_\pi(q^2).
\ee
In the model under investigation it receives two contributions
\be
G_\pi(q^2)=\frac1{g_V} \frac{1+2\beta}{1+\beta}\sum_n\frac{\delta_{n,0}-\delta_{n,1}}{n+1+\mu_V}\left(1-\frac{q^2}{q^2-M^2_V(n)}\right),
\ee
that means that we go beyond the simplest $\rho(770)$-dominance (VMD) approximation. Moreover, a necessary condition is to normalize $G_\pi(0)=1$. That allows us to fix the value of $g_V$:
\be\label{gV_fix}
g_V= \frac{1+2\beta}{(1+\beta)(1+\mu_V)(2+\mu_V)}.
\ee
Hereby, we notice that the introduction of this factor was of the utmost importance to the viability of the model, though we are yet to see its role in the phenomenological fits.
The coupling of the $\rho(770)$ to the pions is then given by
\be
g_{\rho, \pi,\pi}=\sqrt\frac{24\pi^2}{N_c}(1+\mu_V)(2+\mu_V).
\ee
The final expression for the pion FF is
\bea
G_\pi(q^2)&=&1-\frac1{g_V} \sum_n \frac{q^2 F_V(n)}{M^2_V(n)} \frac{g_{\rho_n, \pi,\pi}}{q^2-M^2_V(n)}\\&=&1-\frac{q^2}{q^2-M^2_V(0)}+\frac{q^2M_V^2(0)}{(q^2-M^2_V(0))(q^2-M^2_V(1))},\label{pionFF}
\eea
and its plot can be seen in Fig.~\ref{PionFFGraph}. There we also include as a marker the simplest case of the $\rho(770)$ dominated form factor, it provides a good interpolation in the $Q^2\lesssim1$~GeV$^2$ region but fails at higher energies. The more conventional holographic models predict the pion FF above the VMD result at $Q^2\gtrsim1$~GeV$^2$ as is shown in a summary of HW and SW results in \cite{Kwee2008}, the modified-dilaton SW of \cite{Gherghetta2009} shows a slight improvement, and some other modifications \cite{Sui2010,Fang2016} may bring it closer to but not below the VMD shape. A characteristic feature of our model is that it makes a prediction beyond the VMD result, and that brings it much closer to the experimental points in the most studied region $Q^2\lesssim3$~GeV$^2$. We only find an example of similar behavior achieved in the SW model with an additional quartic term in the scalar potential and a specific and rather complicated form of the scalar vev (model IIb of \cite{Sui2010}). It is also obvious from Fig.~\ref{PionFFGraph} that a higher $\rho$ mass gives a better prediction. The sensitivity to the variations in the $\rho'$ mass is rather negligible. The notion of the value $m_\rho \simeq 1$~GeV originates in an assumption of the ground state positioned on the linear trajectory of the higher radial excitations and does not appear much irrelevant in a holographic construction based on the reproduction of the linear Regge trajectories. We will come back to this option in Section~\ref{Experiment}.

The large-$Q^2$ asymptotics of Eq.~(\ref{pionFF}) is
\be
Q^4 G_\pi(Q^2\rightarrow \infty)=M^2_V(0)(M^2_V(0)+M^2_V(1))+\mathcal{O}\left(\frac1{Q^2}\right)\simeq  1.65~\text{GeV}^2,
\ee
and that is not in accordance with the perturbative QCD expectation of $1/Q^2$-like behavior \cite{PETERLEPAGE1979359}. This is not really surprising after the discrepancies we have seen in the large-$Q^2$ behavior of the two-point functions.

At small $q^2$  we obtain
\be
G_\pi(q^2)=1+q^2\frac1{g_V}\sum_n \frac{F_V(n) g_{\rho_n, \pi,\pi} }{M^4_V(n)} +\mathcal O(q^4)=
1+q^2\left(\frac{1}{M^2_V(0)}+\frac{1}{M^2_V(1)}\right)
\ee
The coefficient at  $q^2$ is associated with the pion charge radius and a chiral coefficient $L_9$
\bea
&2L_9/f_{\pi}^2=\frac16 \langle r^2\rangle^{\pi^\pm}=\frac{1}{M^2_V(0)}+\frac{1}{M^2_V(1)},\\
&L_9=\frac{f_{\pi}^2}{8\kappa^2}\frac{3+2\mu_V}{(1+\mu_V)(2+\mu_V)}.
\eea
Experimentally deduced values of these observables are $L_9=(6.9\pm0.7) \cdot10^{-3}$ \cite{Pich1995} and $r_\pi=\sqrt{\langle r^2\rangle^{\pi^\pm}}=0.659\pm0.004$~fm \cite{PDG2018}.

The $\rho_n a_{1 n_1}\pi_{n_2}$ coupling can also be found from the $5D$ Lagrangian, 
\be
g_{\rho_n, a_{1 n_1},\pi_{n_2}}=\frac {4R}{k_s}\int dze^{-\Phi(\kappa^2z^2)}\frac{1}{z^3}\frac{f(z)R(f(z)+b(z))R\cdot\beta}{\chi_\pi(1+\beta)} D_n(z) C_{n_1}(z)\pi_{n_2}(z). 
\ee
For $n_1=n_2=0$ we calculate
\be
g_{\rho_n, a_1,\pi} =4\kappa \sqrt{\frac{\mu_V}{1+\beta}}\sqrt{\frac{2g_5^2}{R(1+n)}}\left(\delta_{n,0}-\delta_{n,1}\right),
\ee
and using the value of $g_V$ from Eq.~(\ref{gV_fix}), the coupling between the three ground states is
\be
g_{\rho, a_1,\pi} =4\kappa \sqrt{\frac{\mu_V}{1+\beta}}\sqrt{\frac{24\pi^2}{N_c}}\frac{(1+\beta)(1+\mu_V)(2+\mu_V)}{1+2\beta}.
\ee
 
\begin{figure}
\includegraphics[scale=0.2]{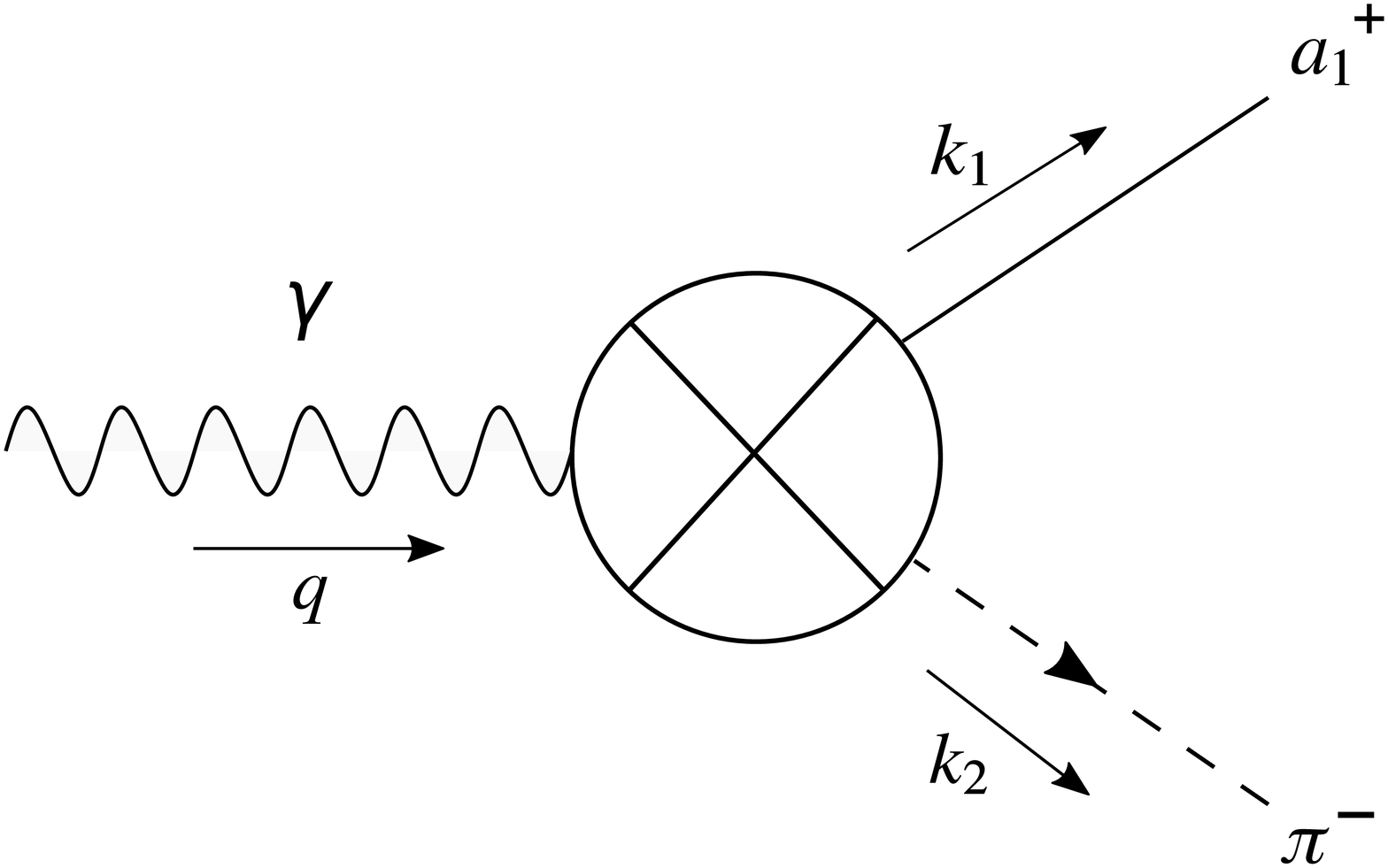}
\caption{Axial form factor. The effective coupling $\bigotimes$ comprises the direct contribution and the one mediated through the $\rho$ mesons.}\label{AxialFigure}
\end{figure}
 The axial form factor as defined by the diagram in Fig.~\ref{AxialFigure} is given by
\bea
G_{a_1}(q^2)=\frac{2\kappa}{g_V} \sqrt\frac{\mu_V}{1+\beta}\sum_n\frac{\delta_{n,0}-\delta_{n,1}}{n+1+\mu_V}\left(1-\frac{q^2}{q^2-M^2_V(n)}\right)=\\
=\frac{2\kappa\sqrt{\mu_V(1+\beta)}}{1+2\beta} \left[1-\frac{q^2}{q^2-M^2_V(0)}+\frac{q^2M_V^2(0)}{(q^2-M^2_V(0))(q^2-M^2_V(1))}\right].
\label{AxialFF}
\eea
Once the model parameters are fixed we can determine from the $q^2$ independent part of this expression the direct coupling in the $a_1\rightarrow \pi\gamma$ process. 
Many holographic models predict zero value for this decay: either due to the absence of the direct term \cite{daRold_2005,Hirn2005} or due to the exact cancellation of $\rho$ and $\rho'$ contributions \cite{PhysRevD.69.065020}.
We consider the fit to this and other observables in the next section.

\section{Fitting the observables}\label{Experiment}
With the QCD parameters fixed, $N_c=3,\ \alpha=\frac1{137}$,
we have three major model parameters $\kappa,\ \mu_V,\ \beta$, and a free parameter $\mu_H$, that is mostly used to set $a_0$ mass to the experimental value (if we neglect the chiral condensate prediction from the scalar two-point function). The parameters $g_5$ and $k_s$ are set standardly by Eq.~(\ref{g_VS_matching}).

Let us resume the estimates we have acquired.
Defined from the one-point function are the constants $F_\rho,\ F_{a_1},\ f_\pi$ in the spin one sector, and $F_s,\ F_\pi$ in the spin zero sector.
The lepton $\rho$ decay,
\be
\Gamma_{\rho\rightarrow e^+e^-}=\frac{4\pi \alpha^2 F_\rho^2}{3m^3_{\rho}}= 7.04\pm0.06\ \text{KeV},
\ee
provides a high precision $F_\rho|_{exp}=0.12124\pm0.00002$~GeV$^2$ \cite{PDG2018}. This predicts $\kappa=519$~MeV. 
For the value of $F_{a_1}$, we can refer to a theoretical (and extra-dimension as well) work \cite{PhysRevD.69.065020}, where they estimate $F_{a_1}=0.26$~GeV$^2$. In addition, they cite (with a proper normalization coinciding with ours) an experimental result \cite{PhysRevD.39.1357}, $0.177\pm0.014$~GeV$^2$, and a lattice one \cite{PhysRevLett.74.4596}, $0.21\pm0.02$~GeV$^2$. 
As was already mentioned, our model implies $F_\rho=F_{a_1}$. However, it is shown further that the estimation achieved in some fits lies in between  $F_\rho|_{exp}$ and $F_{a_1}|_{exp}$. It is a fair result for our holographic setup. Moreover, it is distinct from other approaches. For instance, the usual SW model predicts: 
$F_\rho|_{SW}=0.07\ \text{GeV}^2,\ F_{a_1}|_{SW}=0.31\ \text{GeV}^2$.
Similarly, in a deformed AdS$_5$ background of Ref.~\cite{Ghoroku2006} they find $F_{a_1}$ somewhat higher and $F_\rho$ lower than their experimental values.
The leptonic decay of $\rho'$ is not as widely discussed, but we can predict, with the model's $F_{\rho'}=\sqrt2 F_\rho$, the following decay rate: $\Gamma_{\rho'\rightarrow e^+e^-}=2 \div 3$~KeV in coincidence, for instance, with \cite{Afonin2004}.

To evaluate $f_\pi$, we will use Eq.~(\ref{pionDecay}) with just the first term in this generally diverging sum. This concession brings the model prediction close to the experimental value.

In the scalar sector, we have some information from Ref.~\cite{Gokalp2001}: 
$F_s=0.21\pm0.05\ \text{GeV}^2$. Matching this result requires $\kappa=734$~MeV. Thus, we already see that there is no perfect choice for the $\kappa$ to satisfy both the scalar and vector observables.
The value of $F_\pi$ depends on the estimation of the chiral condensate $\langle q\bar q\rangle =-(235 \div242$~MeV$)^3$\cite{Colangelo2001, Jamin2002} and hence $F_\pi=0.14\div0.15$~GeV$^2$. 
In the relation between $F_s$ and $F_\pi$ the holographic model mimics the situation of $F_\rho$ and $F_{a_1}$ for $\beta=-2$. For a general $\beta$, however, they can be different as the phenomenology would suggest. Thus, on the one hand we can control the difference as opposed to the spin one case. But on the other, we face the fact that for large-$Q^2$, they do not coincide unless $\beta=-2$, which runs against the presumed chiral symmetry restoration at high energies.

In the two-point functions there appear all the phenomenological masses and the aforementioned decay constants. As well, there are the low-energy observable $L_{10}$ and the controversially defined estimations for the condensates  $\langle q \bar q\rangle$ and $\langle G^2\rangle$. 
At the same time, due to the general discordance in definitions related to the large-$Q^2$ limit, we find it instructive to evaluate separately $f_{\pi}$ and $F_\pi$
and use as an independent check the chiral limit condition $f_{\pi} F_\pi=-\langle q\bar q\rangle$. 

There are several decay rates defined by the triple couplings.
The $\rho$ decay exists in our model for $n=0$ and $1$:
\be
\Gamma_{\rho_n\rightarrow \pi^+\pi^-}=
\frac{(m^2_{\rho_n}-4m^2_\pi)^{3/2}}{48\pi m^2_{\rho_n}}\cdot\left(g_{\rho_n, \pi,\pi}+e^2\frac1{g_V}\frac{F_V(n)}{m^2_{\rho_n}}\right)^2.
\ee
We are mostly interested in the experimental result for the ground state $\rho$, $\Gamma_{\rho(770)\rightarrow \pi^+\pi^-}= 147.5\pm0.8\ \text{MeV}$~\cite{PDG2018}. The processes $\rho^+\rightarrow \pi^0\pi^+$ and $\rho^-\rightarrow \pi^0\pi^-$ receive no electromagnetic contribution and it is also measured that
$\Gamma(\rho(770)^0)-\Gamma(\rho(770)^\pm)=0.3\pm1.3\ \text{MeV}$ \cite{PDG2018}.
We also remark that the ratio of the leptonic $\rho$ decay to the pion one has a separate estimation,
$\frac{\Gamma_{\rho\rightarrow e^+e^-}}{\Gamma_{\rho\rightarrow \pi^+\pi^-}}= (0.40\pm0.05)\cdot10^{-4}$ \cite{PDG2018}.

The $a_1\rightarrow \rho\pi$ decay is studied experimentally in $e^+e^-\rightarrow\tau^+\tau^-$ or $\tau^- \rightarrow\pi^-\pi^0\pi^0\nu_\tau$ processes, and the partial width at the tree level is given by the following expression: 
\be
\Gamma_{a_1\rightarrow \rho\pi}=\frac{\sqrt{(m^2_a-(m_\rho+m_\pi)^2)(m^2_a-(m_\rho-m_\pi)^2)}}{48\pi m_a^3} \left(2+\frac{(m^2_a+m^2_\rho-m^2_\pi)^2}{4m_a^2m_\rho^2}\right) g_{\rho, a_{1},\pi}^2.
\ee
Experimentally, it is known that for the case of $a_1\rightarrow (\rho\pi)_\text{S-wave}$ it takes $60.19\%$ of the full decay width, and hence, the value should belong to the area $150 \div360$~MeV.

The pion FF profile was already discussed in detail. From there we use the parameter related to the pion charge radius $r_\pi=\sqrt{\langle r^2\rangle^{\pi^\pm}}$ in the further fittings.
The expression of the axial FF in Eq.~(\ref{AxialFF}) allows us to estimate the decay rate $a_1\rightarrow \pi\gamma$,
\be
\Gamma_{a_1\rightarrow \pi\gamma}=
\frac\alpha4\frac{m^2_a-m^2_\pi}{ m_a^3}G_{a_1}^2(m_a^2/4)=640\pm 246\ \text{KeV}.
\ee
This PDG quoted experimental value is given in Ref.~\cite{Zielinski1984}. It is also mentioned there that the radiative partial decay estimation is sensitive to the assumed $a_1$ resonance mass and the total width, and they use the parameters standard for their time (35 years ago).

\begin{table}
\begin{tabular}{|*{6}{c|}}  
	\hline
 & & RMS fit& $\chi^2$ fit& $\chi^2$ partial A& $\chi^2$ partial B \\
\cline{3-6}
& & $\kappa=532\ \text{MeV}$& $\kappa=520\ \text{MeV}$&
$\kappa=596\ \text{MeV}$& $\kappa=583\ \text{MeV}$\\
Observable& Experiment & $\beta=-2.12$& $\beta=-1.85$&
$\beta=-5.44$& $\beta=-1.63$ \\
& & $\mu_V=-0.50$& $\mu_V=-0.48$& $\mu_V=-0.58$& $\mu_V=-0.46$\\
& & $\mu_H=-0.65$&$\mu_H=-0.61$& $\mu_H=-0.82$ &$\mu_H=-0.79$ \\
	 \hline
	$m_\rho$ & $775.26\pm0.25$ MeV  & $\bf 751.4$&$\bf 753.5$& $\bf 775.2$&$855.9$ \\
    $m_{\rho'}$ & $1465\pm25$ MeV  & $1303.2$&$1283.8$& $1421.6$&$1445.8$ \\
    $m_{a_1}$ & $1230\pm40$ MeV  & $\bf 912.6$&$\bf 919.2$& $\bf 866.8$&$\bf 1056.3$ \\
    $m_{a_1'}$ & $1654\pm19$ MeV  & $1402.4$ &$1387.6$& $1473.5$&$1572.8$ \\
    $m_{\pi'}$ & $1300\pm100$ MeV  & $\bf 1064.8$&$\bf 1039.5$& $\bf 1191.6$&$1165.2$ \\
    $m_{a_0}$ & $980\pm20$ MeV  & $\bf 980$&$\bf 980$& $\bf 980$&$\bf 980$ \\
	$\rho\rightarrow e^+e^-$ & $7.04\pm0.06$ KeV  & $\bf 8.56$&$\bf 7.72$& $12.23$&$8.31$ \\
	$\rho\rightarrow \pi^+\pi^-$ & $147.5\pm0.8$ MeV  & $\bf 219.0$&$\bf 253.4$& $\bf 147.3$&$309.2$ \\
	$\Gamma(\rho^0)-\Gamma(\rho^\pm)$ & $0.3\pm1.3$ MeV  & $1.37$&$\bf 1.40$& $\bf 1.34$&$\bf1.60$ \\
    $\frac{\Gamma(\rho\rightarrow ee)}{\Gamma(\rho\rightarrow \pi\pi)}\cdot10^{4}$ &      $0.40\pm0.05$  & $\bf 0.39$&$\bf 0.30$& $0.83$&$\bf 0.27$ \\
    $a_1\rightarrow \pi\gamma$ & $640\pm246$ KeV  & $\bf 396$&$\bf 396$& $\bf 202$&$\bf 463$ \\
    $a_1\rightarrow \pi\rho$ & $252\pm105$ MeV  & $\bf 75.9$&$\bf 87.3$& $\bf 19.7$&$\bf 110.3$ \\
    $r_\pi$ & $0.659\pm0.04$ fm  & $\bf 0.742$&$\bf 0.744$& $\bf 0.710$&$\bf 0.656$ \\
    $L_{10}\cdot10^{3}$ & $-(5.5\pm0.7)$ & $\bf -8.4$&$\bf -7.8$& $\bf -7.2$&$\bf -7.8$ \\
    $f_\pi$ & $92.07\pm1.2$ MeV  & $\bf 96.4$&$\bf 92.5$& $\bf 92.2$&$\bf 104.6$ \\
    $F_\rho$ & $0.121237(16)$ GeV$^2$  & $\bf 0.1276$&$\bf 0.1216$& $0.1598$&$0.1528$ \\
    $F_s$ & $0.21\pm0.05$ GeV$^2$  & $\bf 0.156$&$\bf 0.149$& $0.196$&$0.187$ \\
    $F_\pi$ & $0.14\pm0.03$ GeV$^2$  & $\bf 0.152$&$\bf 0.155$& $0.153$&$0.213$ \\
	\hline
\end{tabular}
\caption{Global fits. The quantities that were fitted are given in a bold script.}
\label{Table_Global}
\end{table}

Next, we investigate several options to fix the model parameters. We would like to begin with $\beta$ as a free parameter, thus giving priority to phenomenology over the large-$Q^2$ asymptotics of $\Pi_{\pi}(q^2)$.
First, we can make a global fit to the highlighted observables. In holography to get the best fit one often minimizes the RMS error, defined as 
$\varepsilon_{RMS}=\left(\sum\limits_i\frac{(\delta O_i/O_i)^2}{n_{obs}-n_{par}}\right)^{1/2}$, where $O_i$ is an experimental value of an observable, and $\delta O_i$ is a difference between theoretical and experimental expressions. Naturally, this way the experimental errors are not taken into account at all. But the number $\varepsilon_{RMS}$ still communicates the relative precision of the fit and is used to assess the experimental validity of the model as a whole. Though holographic methods do not claim high accuracy and the experimental precision of some of the discussed observables is impossible to reach, we  believe that the more conventional $\chi^2$ method could also be used to provide some extra insight. Thus in Table \ref{Table_Global} we present both approaches. 

Some comments are in order. For the RMS minimization we have omitted the $\Gamma(\rho(770)^0)-\Gamma(\rho(770)^\pm)$ estimation because in this particular situation the error bars, being higher than the mean value, turn out to be particularly important. The inclusion of this observable affects the fit as a whole to the worse, and the model parameters lie in a very different region from any other fit. In this global fit of $15$ observables with $4$ parameters we get the best fit with $\varepsilon_{RMS}=36\ \%$, and we consider it a rather good out-turn.

In the $\chi^2$ minimization\footnote{Obviously, the values of $\chi^2_n$ are huge. We would like to avoid frightening the reader with such numbers and let him or her stay convinced that holographic models are $\sim 30 \ \%$ accurate in {\it some} sense.} the inclusion of the lepton decay of $\rho$ and $F_\rho$ puts a lot of constraint on the fit. Especially, it seems impossible to achieve simultaneously a good result for both the lepton and the pion $\rho$ decays. The $a_1$ decays are also greatly affected by the matching of the model parameters to the more precisely measured $\rho$-related observables. We try to show to what degree some loosening of the fit affects the predictions. In the ``partial A'' fit the accuracy of $\rho\rightarrow \pi\pi$ rate dominates the fit and the $a_1$ decays receive an even worse description. In the ``partial B'' fit, we include the quantities with somewhat larger error bars.  The most interesting effect there is a tendency for the higher $\rho$ mass (resulting, of course, in a very high pion decay rate, though the coupling itself is moderate $g_{\rho, \pi,\pi}=7.39$). The rates of $a_1\rightarrow \rho\pi$ and $a_1\rightarrow \pi\gamma$ come substantially closer to the experiment, as well as the $a_1$ mass itself. The increase of $F_\rho$ towards $F_{a_1}|_{exp}$ once the lepton decay is out of focus is also evident in both partial fits.

The benefit that $F_s$ gets from the freedom in $\beta$ is not substantial, except perhaps for the ``partial A'' fit. Thus, though we introduced a potential difference between $F_s$ and $F_\pi$, other observables turn out to outweigh this bit of phenomenology.

\begin{wraptable}{R}{0.5\textwidth}
\centering
\vspace{-0.4cm}
 \begin{tabular}{|*{2}{c|}}  
	\hline
&  $\kappa=527\ \text{MeV}$\\
Observable& $\beta=-2$\\
& $\mu_V=\mu_H=-0.5$\\
	 \hline
	$m_\rho$ (MeV)  & $\bf 745.3$ \\
    $m_{\rho'}$  (MeV)  & $1290.9$ \\
    $m_{a_1}$ (MeV)  & $\bf 912.8$\\
    $m_{a_1'}$ (MeV)  & $1394.3$\\
    $m_{\pi'}$ (MeV)  & $\bf 1054.0$\\
    $m_{a_0}$ (MeV) & $\bf 1054.0$ \\
	$\rho\rightarrow e^+e^-$ (KeV) &$\bf 8.43$ \\
	$\rho\rightarrow \pi^+\pi^-$ (MeV) & $\bf 219.5$ \\
	$\Gamma(\rho^0)-\Gamma(\rho^\pm)$ (MeV) & $1.36$ \\
    $\frac{\Gamma(\rho\rightarrow ee)}{\Gamma(\rho\rightarrow \pi\pi)}\cdot10^{4}$& $\bf 0.38$ \\
    $a_1\rightarrow \pi\gamma$ (KeV)& $\bf 413$ \\
    $a_1\rightarrow \pi\rho$ (MeV) & $\bf 80.8$\\
    $r_\pi$ (fm)& $\bf 0.749$ \\
    $L_{10}\cdot10^{3}$ &  $\bf -8.6$\\
    $f_\pi$ (MeV) & $\bf 96.9$\\
    $F_\rho$ (GeV$^2$) &$\bf 0.1250$\\
    $F_s$ (GeV$^2$)&$\bf 0.153$ \\
    $F_\pi$ (GeV$^2$)& $\bf 0.153$ \\
	\hline
\end{tabular}
\caption{Single free parameter ($\kappa$) global fit. In bold are the fitted quantities. This is the best fit with $\varepsilon_{RMS}=32\ \%$.}
\label{Table_SelectedFit}
\vspace{-0.1cm}
\end{wraptable}

Next, we recall the theoretical motivation to implant $\beta=-2$ (coincidental large-$Q^2$ behavior in the scalar sector) and $\mu_H=-1/2$ (related to the choice of the $f(z)$ ansatz).
We would also follow the tentative phenomenological preference for the value of $\mu_V$ to be close to $-1/2$, which can be seen in Table~\ref{Table_Global}.
This allows us to suggest a global fit to the observables with the single remaining free factor -- the original SW scale $\kappa$. Table \ref{Table_SelectedFit} shows the result of such fitting. We have fixed $\beta=-2,$ $\mu_V=\mu_H=-1/2$ and looked for the best fit minimizing the RMS error. It is provided by the value $\kappa=527$~MeV. The relative error $\varepsilon_{RMS}=32\ \%$ is not small, but it still manifests a slightly better agreement than that of a completely free RMS minimization due to the bonus of fitting $15$ observables with just one parameter. 

Using this fit, we can calculate the triple couplings
\be\nn
g_{\rho, \pi,\pi}=6.66, \ g_{\rho, a_1,\pi}=6.28\cdot \kappa = 3.3\ \text{GeV}.
\ee
The experimental quantities (meaning the ones extracted from the decay rates for the experimental values of the interacting particles' masses) are $g_{\rho, \pi,\pi}|_{exp}=5.94, \ g_{\rho, a_1,\pi}|_{exp}=3.9\div6.0\ \text{GeV}$. In light of the standard $g_{\rho, \pi,\pi}|_{SW}=3.33$~\cite{Kwee2008} and $g_{\rho, \pi,\pi}|_{HW}=4.28$ or $5.29$~\cite{HW_2005}, the agreement for the $\rho\pi\pi$ coupling seems to be very good.

Let us also take this fit to calculate the gluon condensate $\langle\frac{\alpha_s}{\pi}G^2\rangle$ from Eqs.~(\ref{V2pt} -- \ref{PS2pt}). The estimate with  the correct sign is achieved only from the axial vector two-point function, $0.020$~GeV$^2$ and the pseudoscalar one, $0.16$~GeV$^2$. The predictions are an order different, but we notice that the former is closer to the Shifman-Vainshtein-Zakharov (SVZ) estimate~\cite{SHIFMAN1979385}, and the latter to the lattice one~\cite{CAMPOSTRINI1989393}. The other two give a negative sign for this particular fit, though, for instance, the expression in the scalar correlator provides $0.016$~GeV$^2$ if $\mu_V=-1/2, \ \mu_H=0$ and in principle can lie in the range of the SVZ estimate. We can also extract the gluon condensate from Eq.~(\ref{resonanceLargeQ2}), where the relevant term in the spin zero case provides $0.13$~GeV$^2$.

Unfortunately, $\mu_V\simeq-1/2$ in the presented fits leads to the too small or even wrong sign value of $\langle q\bar q\rangle$ as defined from $\Pi_{LR}$ in Eq.~(\ref{V_ch_cond}).
However, if we turn to the alternative expressions (\ref{resV_ch_cond}) and assume $N_m=0$ (the VMD limit taken to determine $f_\pi$) the prediction with the fit of Table \ref{Table_SelectedFit} is $4\pi\alpha_s\langle q\bar q\rangle^2=4.3\times 10^{-3}$~GeV$^6$. If the proper term in the scalar correlator (\ref{resS_ch_cond}) is used, we get $7.1\times 10^{-3}$~GeV$^6$. These could be related to the assessment of  Ref.~\cite{NARISON2005223}: $(1.0\pm0.2)\times 10^{-3}$~GeV$^6$.
It is of interest that for the $\ln Q^2$ independent quantity such as $\Pi_{LR}$ its holographic dual with the number of resonance cut-off demonstrates a qualitatively relevant behavior, while the $\varepsilon$ cut-off fails.
At last, estimating $\langle q\bar q\rangle$ as a product of $f_\pi$ and $F_\pi$ we get a rather fair result of $\langle q\bar q\rangle=-(241 \div 244\ \text{MeV})^3$ if the ``partial B'' fit is not taken into account.

Coming back to the interpretation of $f(z)$, we can now estimate the constant factor of Eq.~(\ref{fAnsatz}), tentatively related to the quark mass, $m_q=\sqrt{\frac{k_s}{g_5^2}\frac{\mu_V}{\beta}}\kappa=2\kappa\frac{g_V}{g_S}\sqrt{\frac{\mu_V}{3\beta}}$. 
In the global fits of Table \ref{Table_Global}, $g_V=3.7 \div 4.3$; assuming that $g_S \simeq g_V$, 
 we can get $m_q\sim 220 \div 360$~MeV. Such values can only be related to the constituent quark mass, if any physical counterpart should be looked for at all.
 
\begin{wraptable}{R}{0.56\textwidth}
\vspace{-0.5cm}
\centering
\begin{tabular}{|*{3}{c|}}  
	\hline
 & Physical $\rho$& Heavy $\rho$ \\
\cline{2-3}
&  $\kappa=650\ \text{MeV}$& $\kappa=650\ \text{MeV}$\\
Observable& $\beta=-1.19$& $\beta=-1.35$\\
& $\mu_V=-0.65$& $\mu_V=-0.41$\\
& $\mu_H=-0.93$&$\mu_H=-0.93$ \\
	 \hline
	$m_\rho$ (MeV)  & $\bf 775$&$\bf 1000$ \\
    $m_{\rho'}$  (MeV)  & $1514$&$1640$ \\
    $m_{a_1}$ (MeV)  & $\bf 1230$&$\bf 1230$\\
    $m_{a_1'}$ (MeV)  & $1790$ &$1790$\\
    $m_{\pi'}$ (MeV)  & $\bf 1300$&$\bf 1300$ \\
    $m_{a_0}$ (MeV) & $\bf 980$&$\bf 980$ \\
	$\rho\rightarrow e^+e^-$ (KeV) &$17.3$&$8.1$ \\
	$\rho\rightarrow \pi^+\pi^-$ (MeV) & $94.2$&$464.5$ \\
	$\Gamma(\rho^0)-\Gamma(\rho^\pm)$ (MeV) & $1.28$&$1.94$ \\
    $\frac{\Gamma(\rho\rightarrow ee)}{\Gamma(\rho\rightarrow \pi\pi)}\cdot10^{4}$& $1.84$&$0.17$ \\
    $a_1\rightarrow \pi\gamma$ (MeV)& $1.70$&$0.43$ \\
    $a_1\rightarrow \pi\rho$ (MeV) & $84.4$&$129.9$\\
    $r_\pi$ (fm)& $0.701$&$0.566$ \\
    $L_{10}\cdot10^{3}$ &  $-24.5$&$-6.7$\\
    $f_\pi$ (MeV) & $190.6$&$110.7$\\
    $F_\rho$ (GeV$^2$) &$0.190$& $0.190$\\
    $F_s$ (GeV$^2$)&$0.233$&$0.165$ \\
    $F_\pi$ (GeV$^2$)& $0.409$&$0.325$ \\
	\hline
\end{tabular}
\caption{Particular fits. The model parameters are determined to provide the experimental masses marked as bold.}
\label{Table_FixedMasses}
\vspace{-1.5cm}
\end{wraptable}

Finally, we consider some more particular fits in Table \ref{Table_FixedMasses}, focusing on reproducing the masses of the states. It is a common practice to do so, especially normalizing to the experimental value of $m_\rho$ like in the ``Physical $\rho$'' fit. In the ``Heavy $\rho$'' fit we pursue the idea of a higher $\rho$ mass, that would put it on the radial Regge trajectory defined by the $\rho$ excitations.  The fits' parameters alter enough from those of the previous fits to make sizeable deviations for the values of the observables.
Obviously, the results in Table \ref{Table_FixedMasses} are generally less compatible with experiment. However, we notice that between the two fits the ``Heavy $\rho$'' one is significantly better in predicting the lepton $\rho$ decay, the $a_1$ decays, $L_{10}$ and $f_\pi$. It is naturally worse for the pion $\rho$ decay, and the coupling itself is rather large too, $g_{\rho, \pi,\pi}=8.3$.

 \section{Conclusions}\label{Conclusions}
We have constructed a new holographic model of the two-flavor QCD and have addressed multiple aspects of it. We have described the characteristics of dynamical fields in the scalar and vector sectors corresponding to $\rho,\ a_1,\ a_0$ and $\pi$ mesons, analyzed the two-point functions and the structure of the pion and axial FFs, and calculated several hadronic couplings. 
 
 We questioned several steps in the common model-building strategies and looked for possible generalizations there. At the same time, we required analyticity of our solution that prohibited  overcomplication of the model and even suggested some interrelations between its distinct sectors.
 
The primary framework is that of the Soft Wall model, the simplest one validating the confining properties of QCD in the linearity of the predicted Regge trajectories.
The chiral symmetry breaking occurs as a result of the dual process in the bulk and is subject to the model specifics. Not everything turns up in the QCD-like fashion:  there are massless Goldstones and splitting between the vector and axial vector masses, but the OPE-motivated appearance of the chiral condensate in the two-point functions is not exactly met. One can speculate that introducing a more complicated structure of the scalar vev than that of Eq.~(\ref{fAnsatz}) may fix it. It could be also interesting to make simultaneous modifications of the dilaton profile, providing a way to stay consistent with the EOM (e.g., following the lines of \cite{Gherghetta2009}). However, first, we will lose the analyticity of the solutions, and second, we do not believe that the result will turn out significantly better. Treating the large-$Q^2$ limit of QCD in AdS/QCD is wielding a double-edged sword: on one side there is a near-conformality, but on the other is the sidestep from the strongly coupled regime. 
We cannot suggest any new route; matching the leading logarithms is very useful to establish the holographic couplings in terms of $N_c$, and the inconsistency of the subleading terms is to be tolerated. Moreover, in the presented model, the study of the leading logarithms of $\Pi_s$ and $\Pi_\pi$ allowed us to fix one of the parameters.

We developed a new approach to the description of the pions. They appear separated from the vector fields, though it obliges us to break the local gauge invariance in the bulk. We also introduce a specified scalar potential. Requirements of analyticity, masslessness of the pions and fulfillment of the holographic conditions on the boundary define it completely. Our prediction for the pion FF in the region $Q^2\lesssim3$~GeV$^2$ leads us to assume this new rendition as phenomenologically relevant.

 The parametrization of the model is not quite traditional, because we forsake the use of the quark mass and chiral condensate in the scalar vev, exchanging those for $\beta$, and we introduce new parameters in the $5D$ masses: $\mu_V$ and $\mu_H$. Mixing the theoretically and phenomenologically preferred values of these parameters, we came to a one-parameter fit of Table \ref{Table_SelectedFit} that provides a fair 
 description of the experimental quantities. Generally, we find that the typical SW scale, $\kappa$, can be of order $500\div600$~MeV.

 We believe that the presented model is neither too artificial nor oversimplified. On the phenomenological level, it is certainly more successful than the traditional HW or SW models, while the motivation and assumptions beyond our modifications are easily accessible. 
 
 Among other interesting findings, we would like to mention our proposal to regularize some of the divergent at the boundary quantities via cutting the number of contributing resonances. That is an alternative we have not seen utilized often by other authors. It provides some interesting insight in the OPE-related structures and works genuinely well for the estimation of $f_\pi$.

 \begin{acknowledgments}
We acknowledge financial support from the following grants:  FPA2013-46570-C2-1-P (MINECO), 2014SGR104 (Generalitat de Catalunya), and MDM-2014-0369 (MINECO).

We would like to thank Prof. A.A. Andrianov for the valuable discussions on the fundamental topics of this paper.

\end{acknowledgments}

\bibliography{holo_QCD_database}

\end{document}